\documentclass[twocolumn]{aastex61}
\usepackage[figuresright]{rotating}
\begin{document}

\shorttitle{Twenty New TeV Blazars}

\title{Multi-Epoch VLBA Imaging of Twenty New TeV Blazars: Apparent Jet Speeds}

\author{B.~Glenn~Piner}
\affiliation{Department of Physics and Astronomy, Whittier College,
13406 E. Philadelphia Street, Whittier, CA 90608, USA}
\affiliation{Jet Propulsion Laboratory, California Institute of Technology,
4800 Oak Grove Drive, Pasadena, CA 91106, USA}
\email{gpiner@whittier.edu}

\author{Philip~G.~Edwards}
\affiliation{CSIRO Astronomy and Space Science, Australia Telescope National Facility,
PO Box 76, Epping, NSW 1710, Australia}
\email{Philip.Edwards@csiro.au}

\begin{abstract}
We present 88 multi-epoch Very Long Baseline Array (VLBA) images (most at an observing frequency of 8~GHz), of 
20 TeV blazars, all of the HBL class, that have not been previously studied at multiple epochs on the parsec scale.
From these 20 sources, we analyze the apparent speeds of 43 jet components that are
all detected at four or more epochs. As has been found for other TeV HBLs, the apparent
speeds of these components are relatively slow. About two-thirds of the components have
an apparent speed that is consistent (within 2$\sigma$) with no motion;
and some of these components may be stationary patterns whose apparent speed 
does not relate to the underlying bulk flow speed. In addition, a superluminal tail
to the apparent speed distribution of the TeV HBLs is detected for the first time, with 
eight components in seven sources having a 2$\sigma$ lower limit on the apparent speed exceeding $1c$.
We combine the data from these 20 sources with an additional 18 sources from the literature to 
analyze the complete apparent speed distribution of all 38 TeV HBLs that have been studied with VLBI at multiple epochs.
The highest 2$\sigma$ apparent speed lower limit considering all sources is $3.6c$.
This suggests that bulk Lorentz factors of up to about 4, but probably not much higher, exist in the parsec-scale radio
emitting regions of these sources, consistent with estimates obtained in the radio by other means such as
brightness temperatures. This can be reconciled with the high Lorentz factors estimated from the
high-energy data if the jet has velocity structures consisting of different emission regions with different Lorentz factors. 
In particular, we analyze the current apparent speed data for the TeV HBLs
in the context of a model with a fast central spine and a slower outer layer.
\end{abstract}

\keywords{BL Lacertae objects: general --- galaxies: active ---
galaxies: jets --- radio continuum: galaxies}

\section{Introduction}
\label{intro}

High-frequency peaked BL Lac objects (HBLs) constitute the
largest class of active galactic nuclei detected at energies of
$\sim10^{12}$\,eV (1~TeV) with ground-based gamma-ray telescopes, greatly
outnumbering flat-spectrum radio quasars (FSRQs) and low-frequency peaked BL Lac objects (LBLs).
There is considerable evidence that HBLs possess intrinsically weak jets
resulting from radiatively inefficient accretion modes in low-excitation radio
galaxies (LERGs), and that they are thus a physically distinct class of object from the 
intrinsically more powerful FSRQs and LBLs that result from a standard accretion disk
in high-excitation radio galaxies (HERGs) (e.g., Meyer et al.\ 2011; Giommi et al.\ 2012).

These TeV HBLs sometimes display
dramatic variability properties at TeV energies, such as the 200~s variability
timescale seen for PKS 2155$-$304 (Aharonian et al.\ 2007). 
Although various models have been proposed for such rapid variability
(e.g., Begelman et al.\ 2008; Nalewajko et al.\ 2011; Narayan \& Piran 2012; Barkov et al.\ 2012),
they share the common feature of high bulk Lorentz factors and Doppler factors (up to $\delta\sim100$)
for the gamma-ray emitting plasma in their relativistic jets.
High bulk Lorentz factors and Doppler factors are
also required to model TeV blazar spectral energy distributions
(e.g., Tavecchio et al.\ 2010), particularly when one-zone models are used.

In contrast to the estimates obtained at high energies, observations of the parsec-scale
radio jets of HBLs with VLBI have consistently derived modest
values for the bulk Lorentz factor and Doppler factor.
These observations include the low brightness temperatures of the VLBI cores
(e.g., Lister et al.\ 2011; Piner \& Edwards 2014; Lico et al.\ 2016),
the absence of rapid superluminal motions of the jet components
(e.g., Kharb et al.\ 2008; Piner et al.\ 2010; Tiet et al.\ 2012; Lico et al.\ 2012; Lister et al.\ 2016), and others
including the radio variability, core dominance, and jet morphology
(see Piner et al.\ 2008 for a complete discussion).
This discrepancy between the Doppler factors derived at different wavebands for the HBL blazar class was
dubbed the `Doppler crisis' by Tavecchio (2006).

This conflict between the Lorentz and Doppler factor estimates in different wavebands can be resolved if
the HBL jets possess velocity structures such that these quantities vary along the
jet length or width, such as a jet with a fast central spine and a slower outer
layer (e.g., Ghisellini et al.\ 2005), or a jet where the leading edge of ejected
blobs moves faster (Lyutikov \& Lister 2010). We note that such velocity structures in HBL jets are
also independently required in order to match the properties of HBLs with their putative parent objects
(e.g., Chiaberge et al.\ 2000; Meyer et al.\ 2011; Sbarrato et al.\ 2014).
If such velocity structures exist, then some may be directly observable in VLBI imaging as,
for example, limb-brightening of the jet.

High-energy observations, VLBI observations, jet simulations, and unification studies,
when considered together, offer the best approach of obtaining a consistent physical picture of TeV HBL jets.
However, acquiring ample VLBI data to test this has been challenging; because of the relative
faintness of most of the TeV HBLs in the radio, only the brightest
few have been previously well observed with VLBI.

We have been using the Long Baseline Observatory's Very Long Baseline Array (VLBA)
\footnote{The Long Baseline Observatory is a facility of the National Science Foundation 
operated under cooperative agreement by Associated Universities, Inc.}
to conduct a multi-epoch survey of all TeV HBLs accessible to this telescope for
the past several years to study the jet kinematics. When observed over multiple epochs, the apparent motion of the jet
can be directly measured from the VLBI data.
The apparent speed of a jet component moving with the bulk flow speed is given by the well-known formula:
\begin{equation}
\label{speedeqn}
\beta_\mathrm{app}=\frac{\beta\sin\theta}{1-\beta\cos\theta},
\end{equation}
where $\beta c$ is the intrinsic speed and $\theta$ is the angle of the motion to the line of sight.
The apparent speed as a function of $\theta$ has a maximum of $\beta_\mathrm{app}\approx\Gamma$ at
an angle of $\sin\theta=1/\Gamma$, where $\Gamma$ is the bulk Lorentz factor. It is important to note that in a population
of sources with equal Lorentz factors observed at various angles, the peak
measured apparent speed is therefore approximately the Lorentz factor, and slow apparent
speeds are observed at angles both smaller than and larger than the critical angle.
Also note that patterns in the jet may not move with the bulk flow speed, but may instead be 
stationary or slowly moving (e.g., the Low Pattern Speed (LPS) components discussed by
Lister et al.\ 2009b and Piner et al.\ 2012). Such components may be particularly common in the
jets of HBLs due to an abundance of stationary shocks (Hervet et al.\ 2016, 2017), 
and any moving components may need to be observationally separated from a background
of such stationary pattern speed components.

In Piner \& Edwards (2014), hereafter Paper~I, we published first-epoch VLBA images of 
twenty new TeV HBLs that had not been previously well observed with VLBI. In this paper, we
present multi-epoch images of these same 20 sources, and analyze in particular the apparent
speed results that are obtained from these multi-epoch data.
Other results that can be obtained from a single VLBI epoch, such as VLBI core brightness
temperatures and apparent jet opening angles, have already been discussed in Paper~I using
the first epoch of data for each source, and we would not expect such results
from Paper~I to be significantly changed by the additional epochs presented here.
For example, the median VLBI core brightness temperature at 8.4~GHz for all of the images
presented in this paper is $2\times10^{10}$~K, which is the same as the typical brightness temperature
quoted for the TeV HBLs in Paper~I.

Throughout the paper, we assume cosmological parameters of
$H_{0}=71$ km s$^{-1}$ Mpc$^{-1}$, $\Omega_{m}=0.27$, and $\Omega_{\Lambda}=0.73$.
Although different at the few percent level from current best-fit values,
these values are consistent with our earlier publications on TeV blazars
and allow direct comparison with those publications; changing to the
best-fit values would not significantly affect numerical results. 

\begin{table*}[!t]
\begin{center}
\caption{The TeV HBLs}
\label{selecttab}
\begin{tabular}{l c c | l c c} \tableline \tableline \\[-15pt]
\multicolumn{1}{c}{Source\tablenotemark{a}} & Included\tablenotemark{b} & Reason\tablenotemark{c} & 
\multicolumn{1}{c}{Source\tablenotemark{a}} & Included\tablenotemark{b} & Reason\tablenotemark{c}  \\ \tableline \\[-15pt]
SHBL~J001355.9$-$185406 & Y & ... & RX~J1136.5+6737 & N & 4   \\ [-3pt]
KUV~00311$-$1938        & Y & ... & 1ES~1215+303    & N & 2   \\ [-3pt]
1ES~0033+595            & Y & ... & 1ES~1218+304    & N & 1   \\ [-3pt]
RGB~J0136+391           & Y & ... & MS~1221.8+2452  & Y & ... \\ [-3pt]
RGB~J0152+017           & Y & ... & 1ES~1312$-$423  & N & 3   \\ [-3pt]
1ES~0229+200            & Y & ... & PKS~1424+240    & N & 2   \\ [-3pt]
PKS~0301$-$243          & N & 2   & H~1426+428      & N & 1   \\ [-3pt]
IC~310                  & N & 2   & 1ES~1440+122    & Y & ... \\ [-3pt]
RBS~0413                & Y & ... & PKS~1440$-$389  & N & 3   \\ [-3pt]
1ES~0347$-$121          & Y & ... & PG~1553+113     & N & 1   \\ [-3pt]
1ES~0414+009            & Y & ... & Markarian~501   & N & 1   \\ [-3pt]
PKS~0447$-$439          & N & 3   & H~1722+119      & Y & ... \\ [-3pt]
1ES~0502+675            & Y & ... & 1ES~1727+502    & N & 2   \\ [-3pt]
PKS~0548$-$322          & Y & ... & 1ES~1741+196    & Y & ... \\ [-3pt]
RX~J0648.7+1516         & Y & ... & HESS~J1943+213  & N & 4   \\ [-3pt]
1ES~0647+250            & Y & ... & 1ES~1959+650    & N & 1   \\ [-3pt]
RGB~J0710+591           & Y & ... & PKS~2005$-$489  & N & 3   \\ [-3pt]
1ES~0806+524            & N & 2   & 1ES~2037+521    & N & 4   \\ [-3pt]
RBS~0723                & N & 4   & PKS~2155$-$304  & N & 1   \\ [-3pt]
1RXS~J101015.9$-$311909 & Y & ... & RGB~J2243+203   & N & 4   \\ [-3pt]
1ES~1011+496            & N & 2   & B3~2247+381     & Y & ... \\ [-3pt]
1ES~1101$-$232          & N & 1   & 1ES~2344+514    & N & 1   \\ [-3pt]              
Markarian~421           & N & 1   & H~2356$-$309    & N & 1   \\ [-3pt]
Markarian~180           & N & 1   & \\ \tableline \\ [-20pt]
\end{tabular}
\end{center}
\tablenotetext{a}{Source names are the so-called `Canonical Name' used by TeVCat.}
\tablenotetext{b}{ Whether or not the source is included in the VLBA observations for this paper.}
\tablenotetext{c}{Reason for exclusion: 1:~Monitored in our previous work; 2:~in MOJAVE program with sufficient epochs;
3:~too far south; 4:~detection or confirmation too recent, currently being observed as part of our VLBA program BE073.}
\end{table*}

\section{Observations}
\label{observations}
\subsection{Source Selection}
\label{sourceselection}

\begin{table*}[!t]
\begin{center}
\caption{Observed Sources}
\label{sourcetab}
\begin{tabular}{l c l c l l} \tableline \tableline \\ [-15pt]
& B1950 & & VLBA & \multicolumn{1}{c}{First} & \multicolumn{1}{c}{Last} \\ [-3pt]
\multicolumn{1}{c}{Source} & Name & Redshift & Images & 
\multicolumn{1}{c}{Epoch} & \multicolumn{1}{c}{Epoch} \\ \tableline \\ [-15pt] 
SHBL~J001355.9$-$185406 & 0011$-$191 & 0.095                    & 4 & 2013 Aug 16 & 2015 Nov 21 \\ [-3pt]
KUV~00311$-$1938        & 0031$-$196 & 0.506\tablenotemark{a}   & 4 & 2013 Aug 30 & 2015 Nov 21 \\ [-3pt] 
1ES~0033+595            & 0033+595   & 0.467\tablenotemark{b}   & 5 & 2013 Aug 16 & 2015 Nov 21 \\ [-3pt]
RGB~J0136+391           & 0133+388   & 0.400\tablenotemark{c}   & 4 & 2013 Aug 30 & 2015 Nov 21 \\ [-3pt] 
RGB~J0152+017           & 0150+015   & 0.080                    & 5 & 2010 Jan 01 & 2014 Dec 01 \\ [-3pt]
1ES~0229+200            & 0229+200   & 0.140                    & 5 & 2010 Jan 01 & 2014 Dec 01 \\ [-3pt]
RBS~0413                & 0317+185   & 0.190                    & 5 & 2010 Dec 28 & 2014 Dec 01 \\ [-3pt]
1ES~0347$-$121          & 0347$-$121 & 0.188                    & 5 & 2009 Dec 20 & 2014 Dec 01 \\ [-3pt]
1ES~0414+009            & 0414+009   & 0.287                    & 5 & 2010 Dec 28 & 2015 Apr 30 \\ [-3pt]
1ES~0502+675            & 0502+675   & 0.340\tablenotemark{b}   & 4 & 2013 Sep 19 & 2015 Apr 30 \\ [-3pt]
PKS~0548$-$322          & 0548$-$322 & 0.069                    & 4 & 2013 Sep 19 & 2015 Apr 30 \\ [-3pt]
RX~J0648.7+1516         & 0645+153   & 0.179                    & 4 & 2013 Oct 21 & 2015 Aug 02 \\ [-3pt]
1ES~0647+250            & 0647+251   & 0.450                    & 5 & 2013 Oct 21 & 2015 Aug 02 \\ [-3pt]
RGB~J0710+591           & 0706+592   & 0.125                    & 5 & 2010 Feb 16 & 2015 Apr 30 \\ [-3pt]
1RXS~J101015.9$-$311909 & 1008$-$310 & 0.143                    & 4 & 2013 Oct 24 & 2015 Aug 02 \\ [-3pt]
MS~1221.8+2452          & 1221+248   & 0.218                    & 4 & 2013 Oct 24 & 2015 Aug 02 \\ [-3pt]
1ES~1440+122            & 1440+122   & 0.163                    & 4 & 2013 Dec 23 & 2015 Nov 27 \\ [-3pt]
H~1722+119              & 1722+119   & 0.340\tablenotemark{b,d} & 4 & 2013 Dec 23 & 2015 Nov 27 \\ [-3pt]
1ES~1741+196            & 1741+196   & 0.084\tablenotemark{b}   & 4 & 2013 Dec 23 & 2015 Nov 27 \\ [-3pt]
B3~2247+381             & 2247+381   & 0.119                    & 4 & 2013 Dec 23 & 2015 Nov 27 \\ \tableline \\ [-20pt]
\end{tabular}
\end{center}
\tablenotetext{a}{The redshift is a lower limit from Pita et al.\ (2014).}
\tablenotetext{b}{The redshift has been updated compared to that used in Table~7 of Paper~I.}
\tablenotetext{c}{The redshift is a lower limit from Nilsson et al.\ (2012).}
\tablenotetext{d}{The redshift is from Ahnen et al.\ (2016).}
\end{table*}

The goal of our ongoing VLBA project is to obtain multi-epoch VLBI images of the
complete set of TeV-detected HBLs sufficient to study their parsec-scale
jet kinematics and morphology.
Our complete candidate source list is thus the 47 HBLs listed as detections
in the TeVCat catalog\footnote{http://tevcat.uchicago.edu/} (Wakely \& Horan 2008) as of this writing.
These 47 HBLs are listed in Table~\ref{selecttab}. Note that between Paper~I in
2014 and this paper, the total sample size in Table~\ref{selecttab} has grown by only three sources from
44 to 47 objects, implying that there are probably not many TeV HBLs left to be detected
by the current generation of TeV telescopes, and that this table is likely to gain only a 
small number of additions between now and the start of observations with the Cherenkov
Telescope Array (CTA) (Acharya et al.\ 2013).
From this sample of 47 sources, we then excluded the following sources from imaging for this paper:
\begin{enumerate}\addtolength{\itemsep}{-.50\baselineskip}
\item{Eleven sources reported as TeV detections before 2007
for which we have already published multi-epoch VLBA observations: six 
of these sources are discussed by Piner et al.\ (2010),
and an additional five by Tiet et al.\ (2012).}
\item{Seven sources with sufficient multi-epoch VLBA data in the MOJAVE monitoring program
\footnote{http://www.physics.purdue.edu/astro/MOJAVE/allsources.html}.}
\item{Four sources which are below $-35\arcdeg$ declination, and thus are difficult to image with the VLBA.
First epoch images for three of these sources have been published as part of the TANAMI monitoring program
(Ojha et al.\ 2010; M{\"u}ller et al.\ 2017).}
\item{Five sources which were either detected too recently (2014 or later) to be included in this work,
or for which the HBL nature was only recently conclusively established (HESS~J1943+213; Akiyama et al.\ 2016).
Observations of these five sources are currently ongoing as part of our approved VLBA program BE073.}
\end{enumerate} 
The full TeV HBL sample and these exclusions are shown in tabular form in Table~\ref{selecttab}. 
Note that we do not apply any exclusion based on radio flux density; all sources have a flux
density of at least a few millijanskys, and thus are observable with the VLBA.
The applied exclusions leave 20 TeV HBLs
that were all reported as new detections by the TeV telescopes between 2006 and 2013,
and that had not yet been studied with multi-epoch VLBI imaging by any program.
First-epoch VLBA images of these 20 sources were presented in Paper~I, and here we present
new data from additional epochs for each source. 

Table~\ref{sourcetab} gives the B1950 name and the redshift for the 20 sources
observed for this paper, along with the number of VLBA images considered
in the analysis. Hereafter we refer to these sources exclusively by their B1950 names
for uniformity. Redshift values are taken from TeVCat unless otherwise indicated in the notes to Table~\ref{sourcetab}.
Four of these twenty redshift values have been updated based on newer data compared to the 
corresponding redshift values given in Table~7 of Paper~I; these updated redshift values are also
indicated in the notes to Table~\ref{sourcetab}.

The median redshift of our sample of 20 sources is $z=0.18$.
At a redshift of 0.18,
an angle of 1~milliarcsecond (mas) corresponds to a physical length of about 3~parsecs, and a proper motion
of 0.1~mas~yr$^{-1}$ corresponds to a projected linear speed of about 1.2~$c$.

\subsection{Details of Observations}
\label{obsdetails}

\begin{table*}[!t]
\begin{center}
\caption{Observation Log}
\label{obstab}
\begin{tabular}{l l c c l l} \tableline \tableline \\[-15pt]
\multicolumn{1}{c}{Date} & \multicolumn{1}{c}{Observation} & Recording & Observing & 
\multicolumn{1}{c}{Excluded} & \multicolumn{1}{c}{Target Sources} \\ [-3pt]
& Code & Rate & Time & \multicolumn{1}{c}{VLBA} & \\ [-3pt]
& & (Mbps) & (hours) & \multicolumn{1}{c}{Antennas\tablenotemark{a}} & \\ \tableline \\[-15pt]
2009 Dec 20 & BE055B  & 256  & 9  & ...   & 0347$-$121, 0548$-$322\tablenotemark{b}                  \\ [-3pt]
2010 Jan 01 & BE055A  & 256  & 12 & ...   & 0150+015, 0229+200                                       \\ [-3pt]
2010 Feb 16 & BE055C  & 256  & 12 & HN    & 0706+592, 1011+496\tablenotemark{b}                      \\ [-3pt]
2010 Dec 28 & BE057A  & 512  & 12 & ...   & 0317+185, 0414+009, 0502+675\tablenotemark{b}            \\ [-3pt]
2013 Aug 16 & S6117D1 & 2048 & 6  & FD,LA & 0011$-$191, 0033+595                                     \\ [-3pt]
2013 Aug 23 & S6117A1 & 2048 & 8  & ...   & 0150+015, 0229+200, 0317+185, 0347$-$121                 \\ [-3pt]
2013 Aug 30 & S6117D2 & 2048 & 6  & ...   & 0031$-$196, 0133+388                                     \\ [-3pt]
2013 Sep 19 & S6117B1 & 2048 & 8  & KP    & 0414+009, 0502+675, 0548$-$322, 0706+592                 \\ [-3pt]
2013 Oct 21 & S6117D3 & 2048 & 6  & LA    & 0645+153, 0647+251                                       \\ [-3pt]
2013 Oct 24 & S6117D4 & 2048 & 6  & FD,LA & 1008$-$310, 1221+248                                     \\ [-3pt]
2013 Dec 23 & S6117D5 & 2048 & 9  & KP,NL & 1440+122, 1722+119, 1741+196, 2247+381                   \\ [-3pt]
2013 Dec 29 & S6117A2 & 2048 & 8  & HN,KP & 0150+015, 0229+200, 0317+185, 0347$-$121                 \\ [-3pt]
2014 Mar 27 & S6117B2 & 2048 & 8  & ...   & 0414+009, 0502+675, 0548$-$322, 0706+592                 \\ [-3pt]
2014 Jun 11 & S6117A3 & 2048 & 8  & ...   & 0150+015, 0229+200, 0317+185, 0347$-$121                 \\ [-3pt]
2014 Aug 21 & S7017B1 & 2048 & 8  & ...   & 0645+153, 0647+251 (8 \& 15~GHz), 1008$-$310, 1221+248   \\ [-3pt]
2014 Sep 16 & S7017A1 & 2048 & 8  & HN    & 0011$-$191, 0031$-$196, 0033+595 (8 \& 15 GHz), 0133+388 \\ [-3pt]
2014 Nov 10 & S6117B3 & 2048 & 8  & ...   & 0414+009, 0502+675, 0548$-$322, 0706+592                 \\ [-3pt]
2014 Dec 01 & S6117A4 & 2048 & 8  & ...   & 0150+015, 0229+200, 0317+185, 0347$-$121                 \\ [-3pt]
2014 Dec 09 & S7017C1 & 2048 & 8  & HN    & 1440+122, 1722+119, 1741+196, 2247+381                   \\ [-3pt]
2015 Feb 18 & S7017B2 & 2048 & 8  & ...   & 0645+153, 0647+251, 1008$-$310, 1221+248                 \\ [-3pt]
2015 Apr 30 & S6117B4 & 2048 & 8  & OV    & 0414+009, 0502+675, 0548$-$322, 0706+592                 \\ [-3pt]
2015 May 24 & S7017A2 & 2048 & 8  & ...   & 0011$-$191, 0031$-$196, 0033+595, 0133+388               \\ [-3pt]
2015 Jun 07 & S7017C2 & 2048 & 8  & ...   & 1440+122, 1722+119, 1741+196, 2247+381                   \\ [-3pt]
2015 Aug 02 & S7017B3 & 2048 & 8  & ...   & 0645+153, 0647+251, 1008$-$310, 1221+248                 \\ [-3pt]
2015 Nov 21 & S7017A3 & 2048 & 8  & MK    & 0011$-$191, 0031$-$196, 0033+595, 0133+388               \\ [-3pt]
2015 Nov 27 & S7017C3 & 2048 & 8  & ...   & 1440+122, 1722+119, 1741+196, 2247+381                   \\ \tableline \\ [-20pt]
\end{tabular}
\end{center}
\tablenotetext{a}{VLBA antennas that did not participate or that were excluded from the imaging 
for that session. FD=Fort Davis, Texas, HN=Hancock, New Hampshire, KP=Kitt Peak, Arizona, LA=Los Alamos, New Mexico,
MK=Mauna Kea, Hawaii, NL=North Liberty, Iowa, OV=Owens Valley, California.}
\tablenotetext{b}{Observation not included in this paper --- see text for details.}
\end{table*}

Details of all of the observing sessions used for this paper are given in Table~\ref{obstab}.
The bulk of the observations come from VLBA experiments S6117 and S7017 during the years
2013 to 2015. These two experiments together observed each of the 20 sources
from Table~\ref{sourcetab} at four epochs separated by about six months, for
a total of 80 images. Each of these images is obtained from an average of about two hours on-source time;
such integration times are required to image the jets of these fainter sources at sufficient dynamic range. 
These observations were made at an observing frequency of 8.4~GHz (4~cm), because this frequency provides the
optimum combination of angular resolution and sensitivity for these sources.
All observations used the full 2~Gbps recording rate of the VLBA, and were
made using the polyphase filterbank (PFB) observing system of the Roach Digital Backend (RDBE), in its
dual-polarization configuration of eight contiguous 32~MHz channels at matching frequencies in each polarization.
Although dual-polarization was recorded, only total intensity (Stokes I) was calibrated and imaged,
because of the expected sub-millijansky level of polarized flux density from most of these sources.
The only source that was consistently observed using phase-referencing was 0347$-$121
(with J0351$-$1153 observed as the calibrator); other sources
were bright enough for direct fringe fitting.

In addition to the 8.4~GHz images described above, we also obtained images
at 15.3~GHz during experiment S7017 of the two sources 0033+595 and 0647+251.
These two sources displayed apparent jet bends exceeding $90\arcdeg$ in their images
in Paper~I, and images at a higher frequency were obtained in order to attempt
identification of the core in these sources from its spectral properties.

In order to extend the measured time baseline for some sources, we also
included some earlier images from 2009 and 2010 recorded during experiments 
BE055 and BE057, and originally published by Piner \& Edwards (2013).
These images are generally of lower sensitivity, being obtained prior to the VLBA sensitivity
upgrade in 2012. Images from these experiments were included
only if they were of a high enough quality and if the source structure was simple
enough that jet features could be unambiguously connected between the 2009-2010 and 2013-2015
observations. We include earlier observations of the six sources 0150+015, 0229+200, 0317+185, 0347$-$121, 0414+009, 
and 0706+592 (some of which have been reprocessed for this paper), 
and have excluded observations of sources 0502+675, 0548$-$322, and 1011+496
obtained during the same experiments. In the case of 1011+496, the exclusion is because
this source acquired sufficient VLBA epochs through the MOJAVE program.
These additions increase the spanned time range of the VLBA monitoring to
five years for some of these sources, although the typical
spanned time range is closer to two years in most cases.

Altogether, these observations yield a final dataset consisting of 88 images of 20 sources obtained
over the years 2009 to 2015, and totaling approximately 200~hours of integration time on the VLBA.
Twenty of these images were previously published in Paper~I, six were published by Piner \& Edwards (2013),
four were published in Piner \& Edwards (2016), and 58 are previously unpublished.

We used the AIPS software package for calibration and fringe-fitting of the correlated visibilities,
and fringes were found at significant SNR to all target sources at all epochs.
A small number of discrepant visibilities were flagged, 
and the final images were produced using CLEAN and self-calibration in
the DIFMAP software package. 
VLBA imaging of sources at these lower flux density levels can be very sensitive
to the self-calibration averaging interval,
and self-calibration will generate spurious point-source structure if the averaging interval
is too short (e.g., Mart{\'{\i}}-Vidal \& Marcaide 2008).
We carefully investigated and selected self-calibration solution intervals for the fainter
sources to make sure that minimal spurious flux density (less than $\sim1$~mJy) 
should be introduced into the images through self-calibration
(see Equations~7 and 8 of Mart{\'{\i}}-Vidal \& Marcaide 2008).
In the section below, all of the images are displayed using natural weighting, 
in order to maximize the dynamic range.

\section{Results}
\label{results}

\subsection{Images}
\label{images}

\begin{table*}[!h]
\begin{center}
\caption{Parameters of the Images}
\label{imtab}
\begin{tabular}{c l c l c c c} \tableline \tableline \\ [-15pt]
Source & \multicolumn{1}{c}{Epoch} & Frequency & \multicolumn{1}{c}{Beam} & Peak Flux 
& $I_{\mathrm{rms}}$\tablenotemark{b} & Ref.\tablenotemark{c} \\ [-3pt]
& & (GHz) & \multicolumn{1}{c}{Parameters\tablenotemark{a}} & Density & (mJy bm$^{-1}$) \\ [-3pt]
& & & & (mJy bm$^{-1}$) \\ \tableline \\ [-15pt]
0011$-$191 & 2013 Aug 16 & 8.4  & 2.15, 0.82, $-$4.1  & 10  & 0.029 & 1   \\ [-3pt]
           & 2014 Sep 16 & 8.4  & 2.35, 1.01, 2.6     & 11  & 0.033 & ... \\ [-3pt]
           & 2015 May 24 & 8.4  & 2.38, 1.01, 0.0     & 11  & 0.027 & ... \\ [-3pt]
           & 2015 Nov 21 & 8.4  & 3.92, 1.20, 13.6    & 9   & 0.027 & ... \\ [-3pt]
0031$-$196 & 2013 Aug 30 & 8.4  & 2.34, 0.93, $-$1.1  & 26  & 0.022 & 1   \\ [-3pt]
           & 2014 Sep 16 & 8.4  & 2.46, 1.06, 7.0     & 21  & 0.029 & ... \\ [-3pt]
           & 2015 May 24 & 8.4  & 2.35, 0.96, 0.3     & 31  & 0.028 & ... \\ [-3pt]
           & 2015 Nov 21 & 8.4  & 3.81, 1.09, 17.6    & 33  & 0.025 & ... \\ [-3pt]
0033+595   & 2013 Aug 16 & 8.4  & 1.61, 0.84, 0.7     & 43  & 0.024 & 1   \\ [-3pt]
           & 2014 Sep 16 & 8.4  & 1.99, 1.29, 14.3    & 49  & 0.026 & ... \\ [-3pt]
           & 2014 Sep 16 & 15.3 & 1.14, 0.75, 19.5    & 37  & 0.045 & ... \\ [-3pt]
           & 2015 May 24 & 8.4  & 1.67, 1.06, $-$28.6 & 50  & 0.022 & ... \\ [-3pt]
           & 2015 Nov 21 & 8.4  & 1.68, 1.37, $-$8.9  & 49  & 0.022 & ... \\ [-3pt]
0133+388   & 2013 Aug 30 & 8.4  & 1.93, 0.93, 13.3    & 35  & 0.020 & 1   \\ [-3pt]
           & 2014 Sep 16 & 8.4  & 2.31, 1.23, 13.7    & 31  & 0.027 & ... \\ [-3pt]
           & 2015 May 24 & 8.4  & 1.80, 1.03, $-$10.2 & 30  & 0.022 & ... \\ [-3pt]
           & 2015 Nov 21 & 8.4  & 1.95, 1.29, 9.7     & 27  & 0.022 & ... \\ [-3pt]
0150+015   & 2010 Jan 01 & 8.4  & 2.29, 0.97, $-$1.4  & 35  & 0.046 & 2   \\ [-3pt]
           & 2013 Aug 23 & 8.4  & 2.12, 0.92, 0.9     & 43  & 0.025 & 1   \\ [-3pt]
           & 2013 Dec 29 & 8.4  & 2.17, 0.99, 2.2     & 45  & 0.036 & ... \\ [-3pt]
           & 2014 Jun 11 & 8.4  & 2.15, 0.97, 1.0     & 40  & 0.026 & ... \\ [-3pt]
           & 2014 Dec 01 & 8.4  & 2.08, 0.98, 6.0     & 48  & 0.026 & 3   \\ [-3pt]
0229+200   & 2010 Jan 01 & 8.4  & 1.92, 1.07, 1.5     & 17  & 0.046 & 2   \\ [-3pt]
           & 2013 Aug 23 & 8.4  & 1.93, 0.94, $-$0.2  & 21  & 0.023 & 1   \\ [-3pt]
           & 2013 Dec 29 & 8.4  & 1.93, 0.96, 1.9     & 18  & 0.023 & ... \\ [-3pt]
           & 2014 Jun 11 & 8.4  & 1.92, 1.01, 0.1     & 17  & 0.022 & ... \\ [-3pt]
           & 2014 Dec 01 & 8.4  & 1.74, 0.97, 0.3     & 17  & 0.025 & 3   \\ [-3pt]
0317+185   & 2010 Dec 28 & 8.4  & 1.88, 1.04, 0.1     & 19  & 0.030 & 2   \\ [-3pt]
           & 2013 Aug 23 & 8.4  & 1.89, 0.94, 1.3     & 18  & 0.025 & 1   \\ [-3pt]
           & 2013 Dec 29 & 8.4  & 1.89, 1.01, 5.0     & 15  & 0.027 & ... \\ [-3pt]
           & 2014 Jun 11 & 8.4  & 1.92, 1.00, 2.4     & 17  & 0.022 & ... \\ [-3pt]
           & 2014 Dec 01 & 8.4  & 1.74, 0.95, 1.0     & 17  & 0.026 & 3   \\ [-3pt]
0347$-$121 & 2009 Dec 20 & 8.4  & 2.36, 0.93, $-$2.8  & 4   & 0.049 & 2   \\ [-3pt]
           & 2013 Aug 23 & 8.4  & 2.25, 0.89, $-$0.9  & 7   & 0.025 & 1   \\ [-3pt]
           & 2013 Dec 29 & 8.4  & 2.11, 0.90, 1.7     & 5   & 0.027 & ... \\ [-3pt]
           & 2014 Jun 11 & 8.4  & 2.24, 0.94, $-$0.1  & 6   & 0.025 & ... \\ [-3pt]
           & 2014 Dec 01 & 8.4  & 2.06, 0.86, 0.8     & 6   & 0.029 & 3   \\ [-3pt]
0414+009   & 2010 Dec 28 & 8.4  & 2.64, 1.10, 0.2     & 24  & 0.055 & 2   \\ [-3pt]
           & 2013 Sep 19 & 8.4  & 2.04, 0.87, $-$1.7  & 35  & 0.022 & 1   \\ [-3pt]
           & 2014 Mar 27 & 8.4  & 2.16, 0.92, 0.8     & 36  & 0.022 & ... \\ [-3pt]
           & 2014 Nov 10 & 8.4  & 2.19, 0.94, $-$1.7  & 40  & 0.022 & ... \\ [-3pt]
           & 2015 Apr 30 & 8.4  & 2.16, 0.99, 1.6     & 37  & 0.023 & ... \\ [-3pt]
0502+675   & 2013 Sep 19 & 8.4  & 1.34, 1.01, 0.5     & 19  & 0.023 & 1   \\ [-3pt]
           & 2014 Mar 27 & 8.4  & 1.55, 1.08, 11.3    & 15  & 0.020 & ... \\ [-3pt]
           & 2014 Nov 10 & 8.4  & 1.38, 1.07, $-$4.6  & 17  & 0.019 & ... \\ [-3pt]
           & 2015 Apr 30 & 8.4  & 1.50, 1.05, $-$23.7 & 20  & 0.023 & ... \\ [-3pt]
\end{tabular}
\end{center}
\end{table*}

\begin{table*}
\begin{center}
{\bf Table 4} (Continued) \\
\begin{tabular}{c l c l c c c} \tableline \tableline \\ [-15pt]
Source & \multicolumn{1}{c}{Epoch} & Frequency & \multicolumn{1}{c}{Beam} & Peak Flux 
& $I_{\mathrm{rms}}$\tablenotemark{b} & Ref.\tablenotemark{c} \\ [-3pt]
& & (GHz) & \multicolumn{1}{c}{Parameters\tablenotemark{a}} & Density & (mJy bm$^{-1}$) \\ [-3pt]
& & & & (mJy bm$^{-1}$) \\ \tableline \\ [-15pt]
0548$-$322 & 2013 Sep 19 & 8.4  & 2.19, 0.84, 1.0     & 20  & 0.062 & 1   \\ [-3pt]
           & 2014 Mar 27 & 8.4  & 2.36, 0.93, 4.0     & 21  & 0.029 & ... \\ [-3pt]
           & 2014 Nov 10 & 8.4  & 2.43, 0.92, $-$0.2  & 24  & 0.034 & ... \\ [-3pt]
           & 2015 Apr 30 & 8.4  & 2.30, 0.86, $-$0.6  & 27  & 0.032 & ... \\ [-3pt]
0645+153   & 2013 Oct 21 & 8.4  & 1.92, 0.86, $-$3.0  & 36  & 0.020 & 1   \\ [-3pt]
           & 2014 Aug 21 & 8.4  & 2.14, 0.98, 2.1     & 23  & 0.025 & ... \\ [-3pt]
           & 2015 Feb 18 & 8.4  & 2.13, 1.06, 1.2     & 22  & 0.022 & ... \\ [-3pt]
           & 2015 Aug 02 & 8.4  & 2.09, 0.96, 1.7     & 25  & 0.023 & ... \\ [-3pt]
0647+251   & 2013 Oct 21 & 8.4  & 1.88, 0.88, $-$4.9  & 43  & 0.018 & 1   \\ [-3pt]
           & 2014 Aug 21 & 8.4  & 2.08, 1.00, 1.4     & 50  & 0.029 & ... \\ [-3pt]
           & 2014 Aug 21 & 15.3 & 1.07, 0.49, $-$4.9  & 40  & 0.048 & ... \\ [-3pt]
           & 2015 Feb 18 & 8.4  & 2.06, 1.04, $-$2.5  & 53  & 0.022 & ... \\ [-3pt]
           & 2015 Aug 02 & 8.4  & 2.04, 0.95, 1.1     & 53  & 0.025 & ... \\ [-3pt]
0706+592   & 2010 Feb 16 & 8.4  & 1.42, 1.10, 1.3     & 28  & 0.038 & 2   \\ [-3pt]
           & 2013 Sep 19 & 8.4  & 1.42, 1.03, 14.5    & 28  & 0.023 & 1   \\ [-3pt]
           & 2014 Mar 27 & 8.4  & 1.67, 1.13, 34.2    & 29  & 0.022 & ... \\ [-3pt]
           & 2014 Nov 10 & 8.4  & 1.46, 1.07, 5.7     & 32  & 0.020 & ... \\ [-3pt]
           & 2015 Apr 30 & 8.4  & 1.47, 1.02, 1.6     & 32  & 0.022 & ... \\ [-3pt]
1008$-$310 & 2013 Oct 24 & 8.4  & 2.20, 0.81, $-$2.7  & 29  & 0.040 & 1   \\ [-3pt]
           & 2014 Aug 21 & 8.4  & 2.56, 0.92, 1.8     & 26  & 0.040 & ... \\ [-3pt]
           & 2015 Feb 18 & 8.4  & 2.69, 1.02, 2.6     & 21  & 0.036 & ... \\ [-3pt]
           & 2015 Aug 02 & 8.4  & 2.53, 1.00, 5.8     & 15  & 0.064 & ... \\ [-3pt]
1221+248   & 2013 Oct 24 & 8.4  & 1.83, 0.84, $-$0.1  & 16  & 0.023 & 1   \\ [-3pt]
           & 2014 Aug 21 & 8.4  & 2.04, 0.88, 4.9     & 17  & 0.026 & ... \\ [-3pt]
           & 2015 Feb 18 & 8.4  & 2.10, 1.05, 2.4     & 16  & 0.021 & ... \\ [-3pt]
           & 2015 Aug 02 & 8.4  & 1.95, 1.02, 12.1    & 15  & 0.022 & ... \\ [-3pt]
1440+122   & 2013 Dec 23 & 8.4  & 1.98, 0.87, $-$7.2  & 18  & 0.025 & 1   \\ [-3pt]
           & 2014 Dec 09 & 8.4  & 2.21, 0.95, $-$4.0  & 19  & 0.020 & ... \\ [-3pt]
           & 2015 Jun 07 & 8.4  & 2.11, 0.92, $-$6.3  & 17  & 0.019 & ... \\ [-3pt]
           & 2015 Nov 27 & 8.4  & 2.05, 0.89, $-$4.3  & 17  & 0.021 & ... \\ [-3pt]
1722+119   & 2013 Dec 23 & 8.4  & 2.04, 0.98, $-$11.8 & 66  & 0.030 & 1   \\ [-3pt]
           & 2014 Dec 09 & 8.4  & 2.12, 1.02, 0.5     & 107 & 0.025 & ... \\ [-3pt]
           & 2015 Jun 07 & 8.4  & 2.06, 0.97, $-$1.8  & 62  & 0.022 & ... \\ [-3pt]
           & 2015 Nov 27 & 8.4  & 1.98, 0.95, $-$1.7  & 62  & 0.024 & ... \\ [-3pt]
1741+196   & 2013 Dec 23 & 8.4  & 1.95, 0.97, $-$12.6 & 98  & 0.030 & 1   \\ [-3pt]
           & 2014 Dec 09 & 8.4  & 2.04, 1.01, 0.2     & 101 & 0.025 & ... \\ [-3pt]
           & 2015 Jun 07 & 8.4  & 1.96, 0.96, $-$2.3  & 109 & 0.023 & ... \\ [-3pt]
           & 2015 Nov 27 & 8.4  & 1.92, 0.94, $-$4.9  & 121 & 0.025 & ... \\ [-3pt]
2247+381   & 2013 Dec 23 & 8.4  & 2.11, 0.81, 2.5     & 42  & 0.029 & 1   \\ [-3pt]
           & 2014 Dec 09 & 8.4  & 2.27, 1.03, 31.4    & 48  & 0.023 & ... \\ [-3pt]
           & 2015 Jun 07 & 8.4  & 2.01, 0.95, 25.8    & 48  & 0.024 & ... \\ [-3pt]
           & 2015 Nov 27 & 8.4  & 1.99, 0.97, 26.9    & 56  & 0.025 & ... \\ \tableline \\ [-20pt]
\end{tabular}
\end{center}
\tablenotetext{a}{FWHM of the major
and minor axes in mas, and position angle of the major axis in degrees; respectively.}
\tablenotetext{b}{rms noise in the total intensity image.}
\tablenotetext{c}{References for previously published images: (1) Piner \& Edwards 2014 (Paper~I); (2) Piner \& Edwards 2013;
(3) Piner \& Edwards 2016}
\end{table*}

\begin{figure*}[!t]
\centering
\includegraphics[scale=0.82,trim={0 0.5cm 3.4cm 0}]{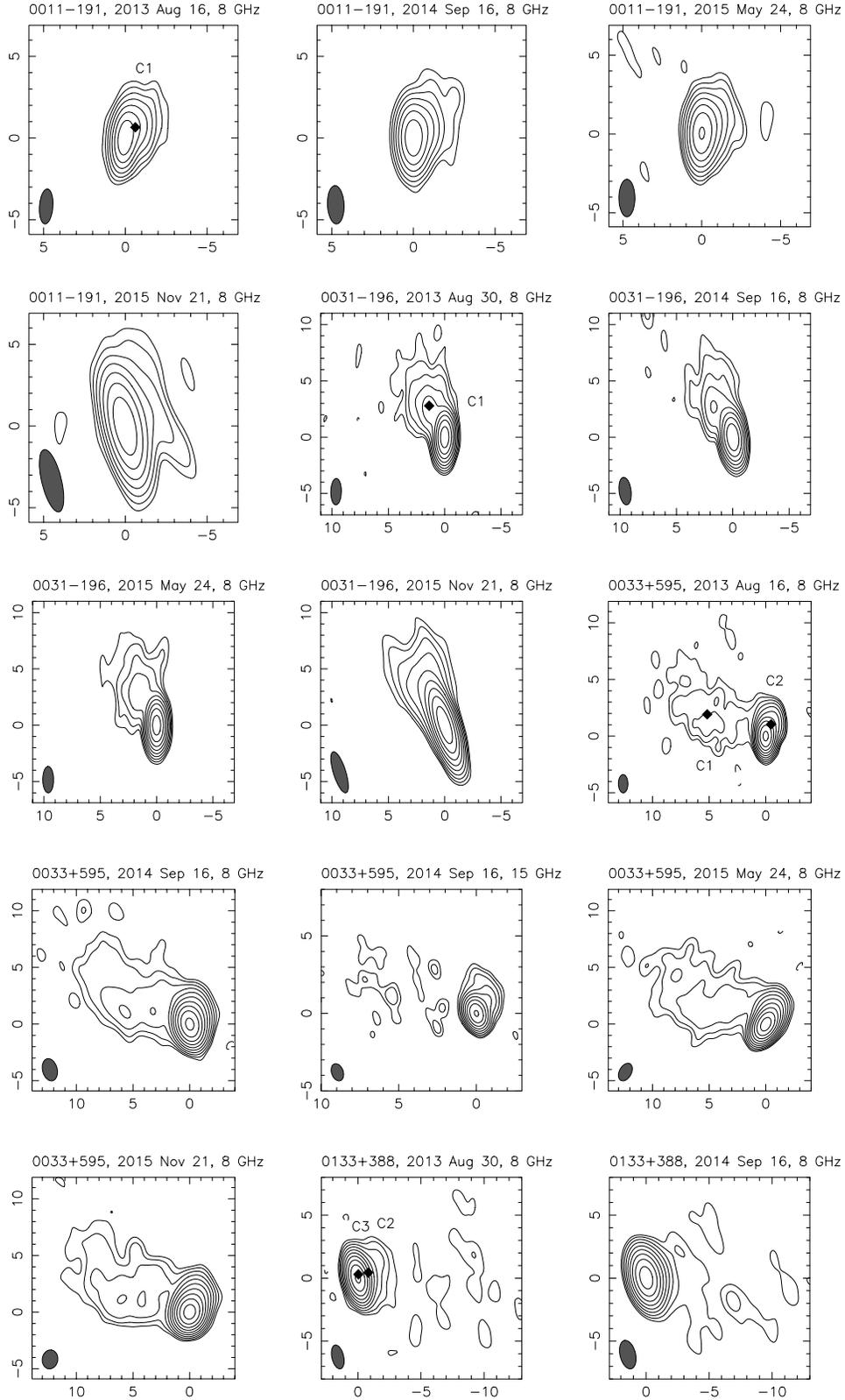}
\vspace{-0.15in}
\caption{VLBA images.
Parameters are in Table~\ref{imtab}. Axes are in mas.
The lowest contour is three times the noise level;
other contours are factors of two higher.
Filled diamonds in first epoch images (from 2013)
indicate locations of fitted Gaussian jet components ---
see text for full description.}
\end{figure*}

\begin{figure*}
\centering
\includegraphics[scale=0.82,trim={0 0.5cm 3.4cm 0}]{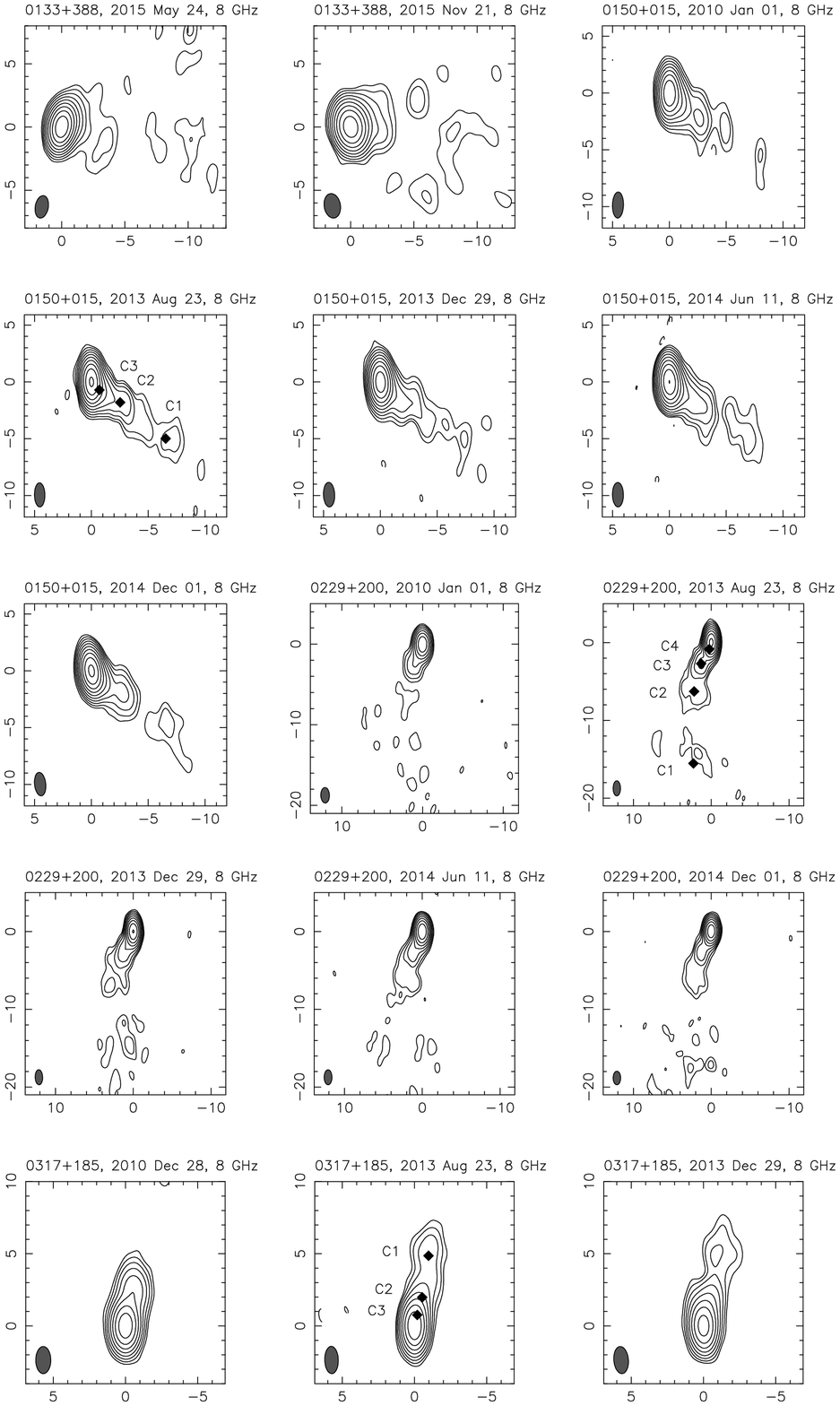}
\\Fig. 1.--—{\em Continued}
\end{figure*}

\begin{figure*}
\centering
\includegraphics[scale=0.82,trim={0 0.5cm 3.4cm 0}]{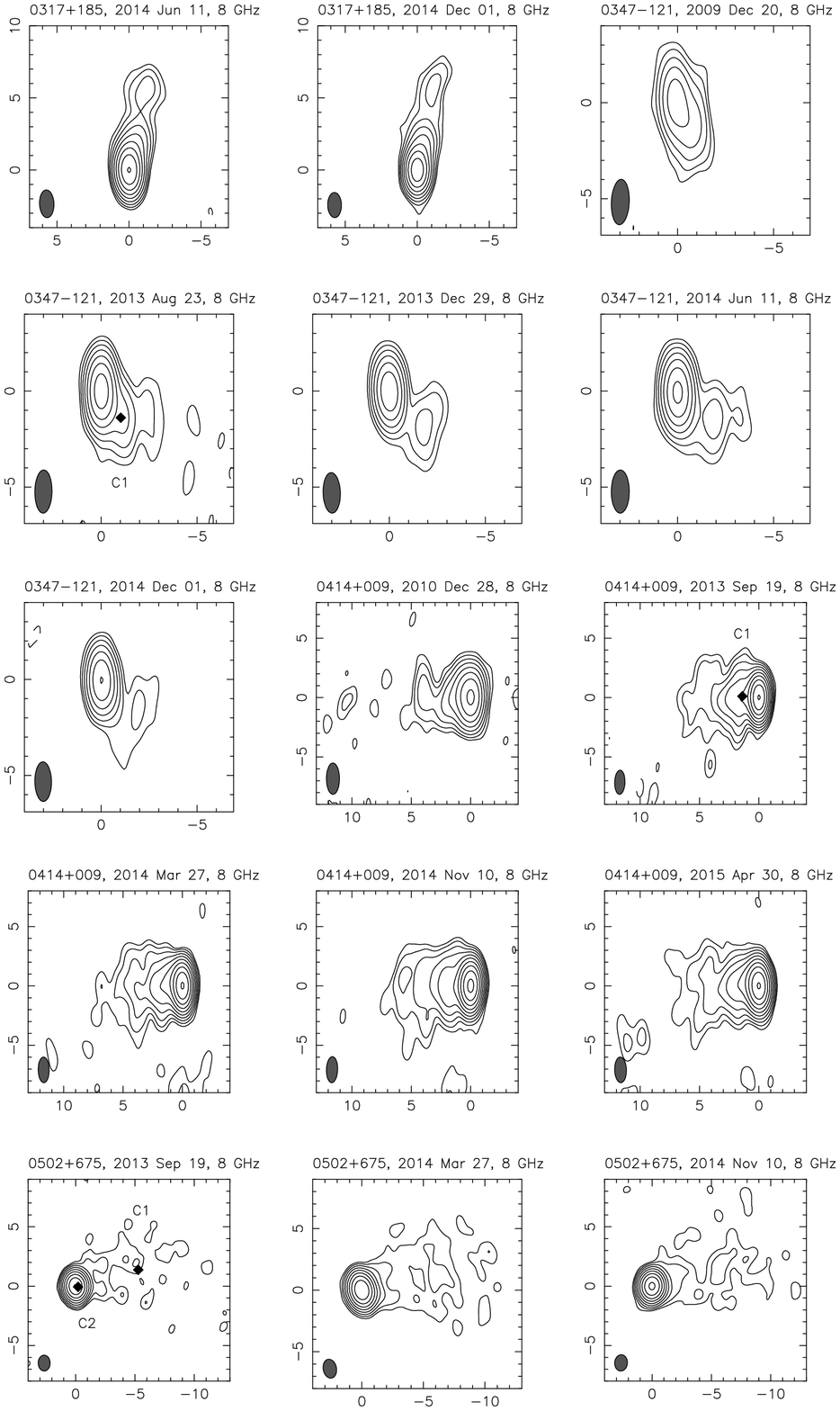}
\\Fig. 1.--—{\em Continued}
\end{figure*}

\begin{figure*}
\centering
\includegraphics[scale=0.82,trim={0 0.5cm 3.4cm 0}]{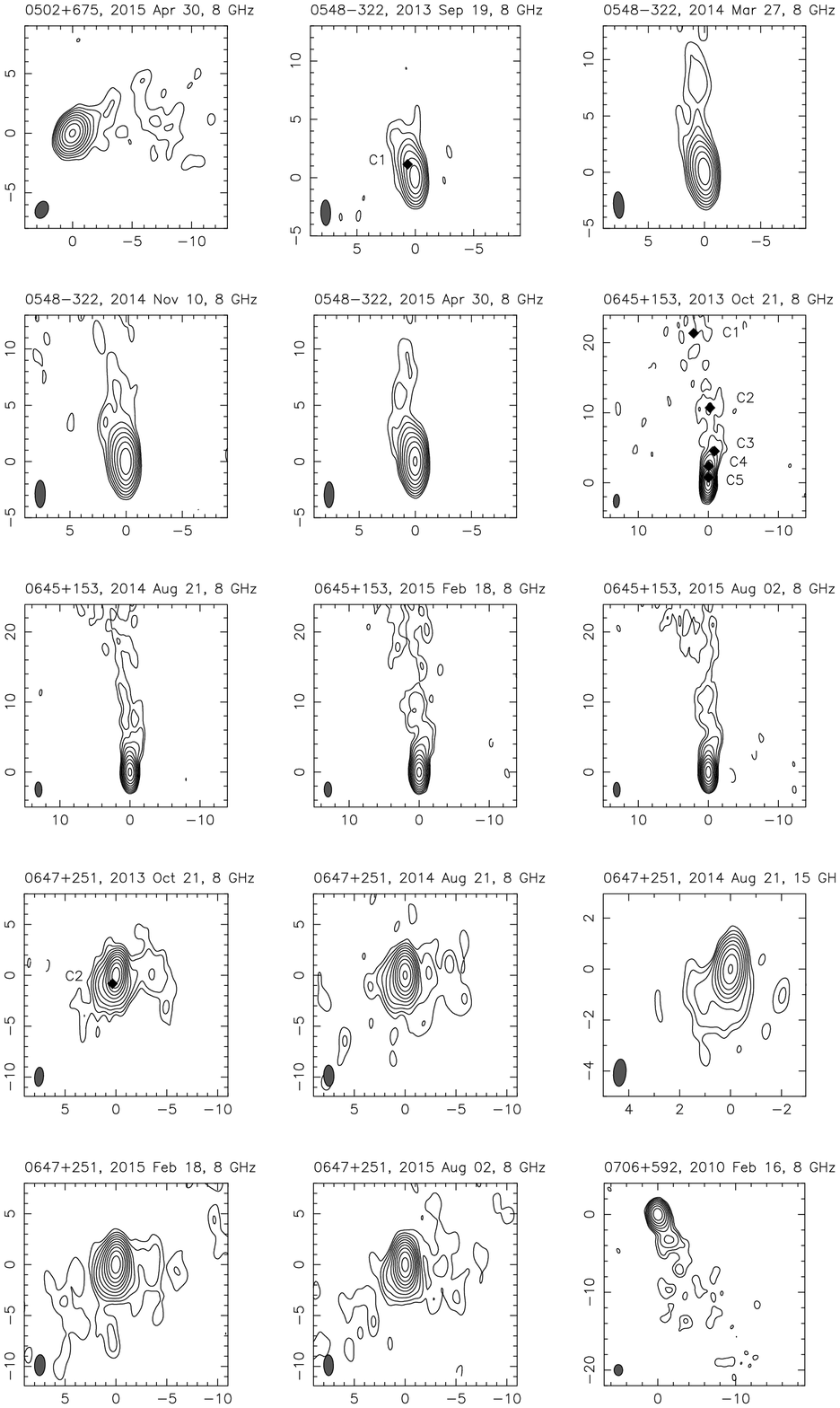}
\\Fig. 1.--—{\em Continued}
\end{figure*}

\begin{figure*}
\centering
\includegraphics[scale=0.82,trim={0 0.5cm 3.4cm 0}]{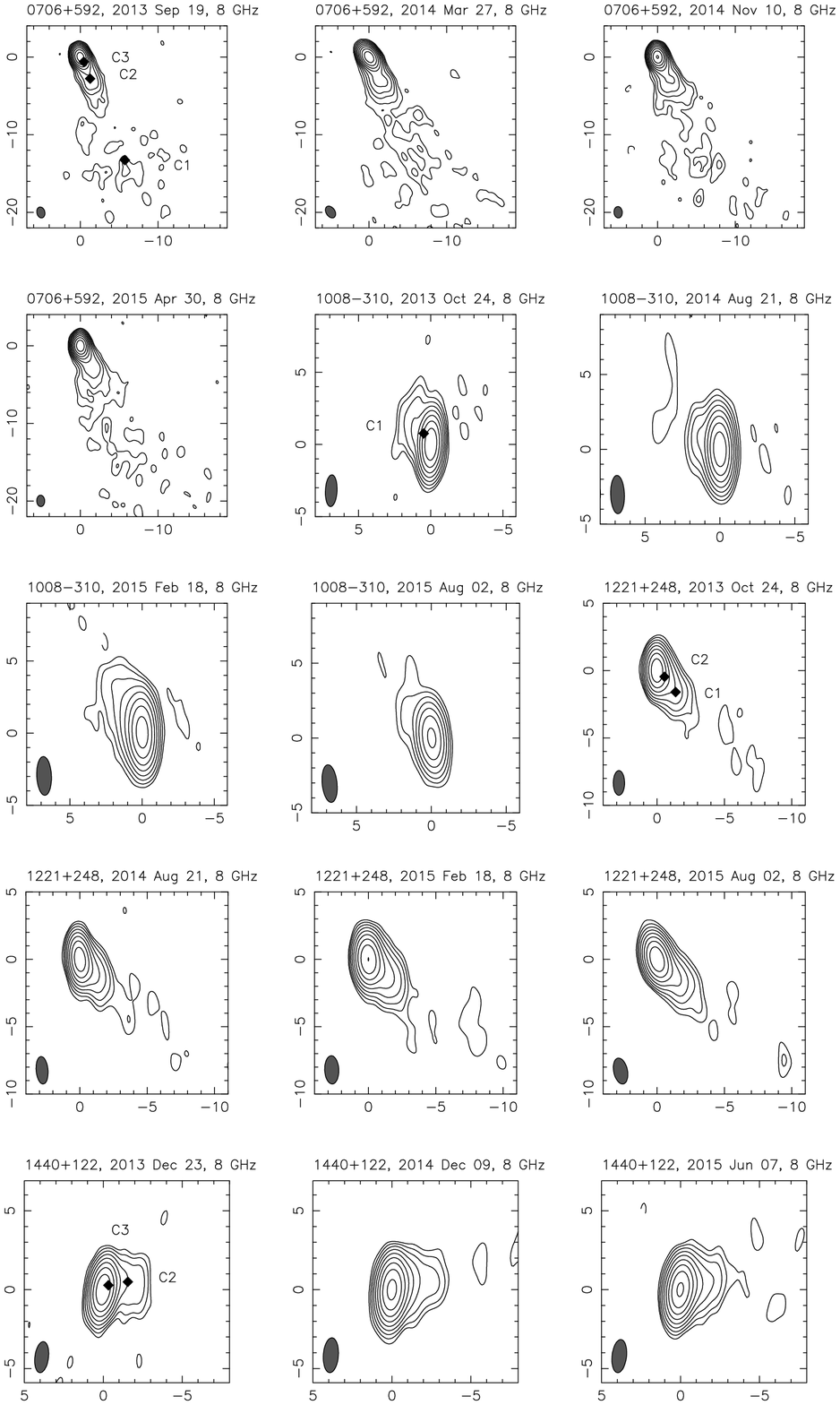}
\\Fig. 1.--—{\em Continued}
\end{figure*}

\begin{figure*}
\centering
\includegraphics[scale=0.82,trim={0 0.5cm 3.4cm 0}]{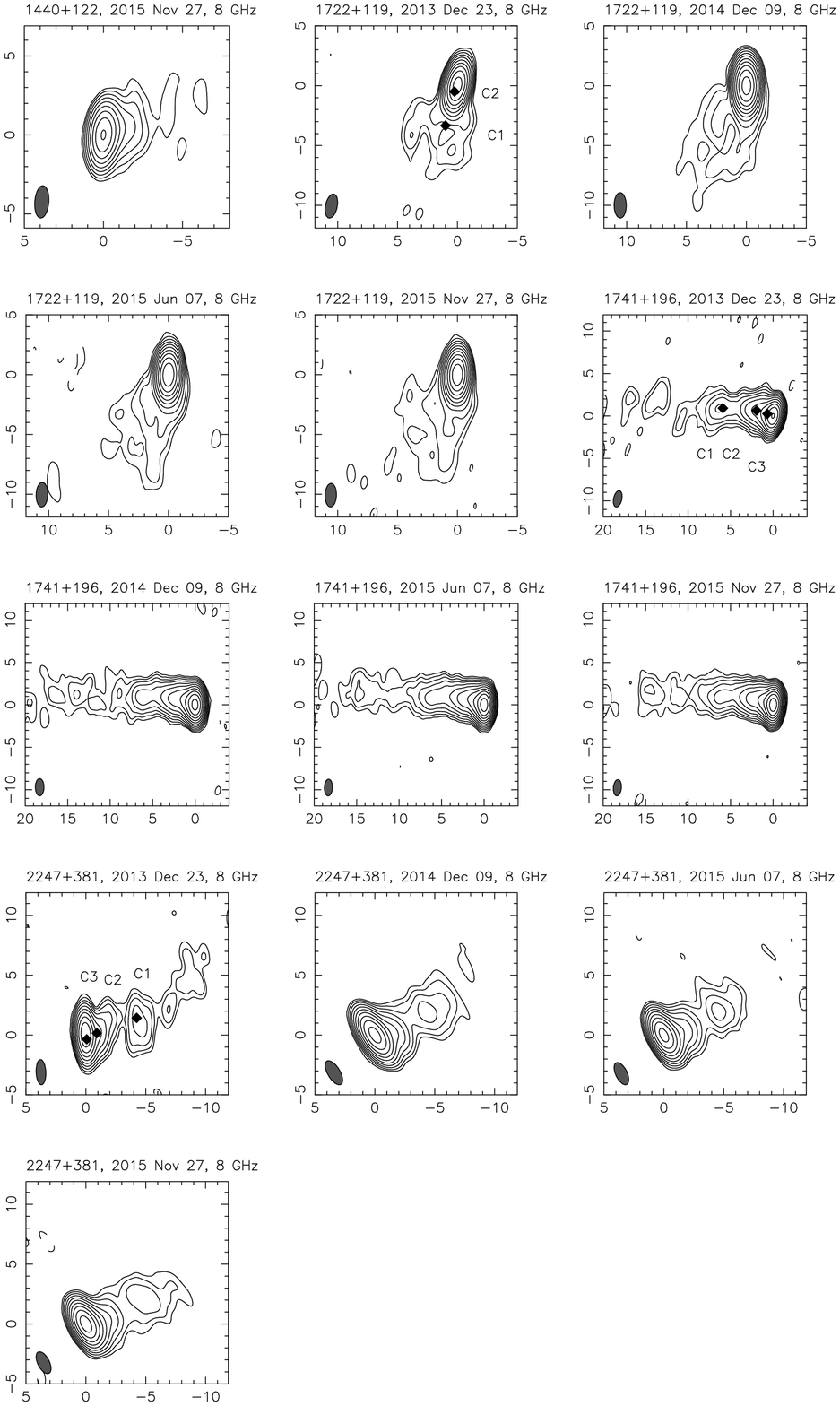}
\\Fig. 1.--—{\em Continued}
\end{figure*}

The 88 VLBA images used for this paper are shown in Figure~1, and the
parameters of these images are tabulated in Table~\ref{imtab}.
The source name, epoch, and observing frequency are listed above each panel in Figure~1.
All sources show a bright, compact component, hereafter identified as the VLBI core,
and they all show additional extended structure that can be modeled by at least one
Gaussian feature in addition to the core (see $\S$~\ref{mfits}).
The fitted locations of the 43 Gaussian components for which apparent speeds are determined
in $\S$~\ref{speeds} are indicated by filled diamonds on the first image from 2013 for each source
(which is the image published in Paper~I, and also typically the one with the highest dynamic range); this is intended
to aid comparison between the images in Figure~1 and the fits in Figure~2 (see $\S$~\ref{speeds}).
The images in Figure~1 do not show the entire CLEANed region for clarity, but are instead zoomed in on
the core and the inner jet region.
Larger scale images plus all associated data files are available at the project web site
\footnote{http://whittierblazars.com}.

Peak flux densities in the images in Figure~1 range from 4 to 121 mJy~bm$^{-1}$
(see Table~\ref{imtab}), with a median peak flux density of 27 mJy~bm$^{-1}$.
The median rms noise level is 0.025 mJy~bm$^{-1}$, which is about the
expected noise level for an approximately two-hour observation at 8.4~GHz
\footnote{https://science.nrao.edu/facilities/vlba/docs/manuals/oss/imag-sens}.
Typical dynamic ranges of the images in Figure~1 are thus about 1000:1, 
which is easily high enough to reveal the parsec-scale jet structure even in the fainter sources.

The general parsec-scale morphology of these sources was described in Paper~I, and
has also been seen in VLBI studies of brighter TeV blazars such as Mrk~421 and Mrk~501 
(Piner et al.\ 1999; Edwards \& Piner 2002; Giroletti et al.\ 2006, 2008).
Most of the sources show a collimated jet a few milliarcseconds long that transitions
to a lower surface brightness, more diffuse jet with a broader opening angle at a few mas from the core.
The structure at tens of milliarcseconds from the core at 8~GHz then appears patchy and filamentary.

In Paper~I we noted that at least two sources (0502+675 and 1722+119) showed a clear limb-brightened
structure at a few mas from the core; once again, this is a property that is familiar from the brighter TeV blazars
(e.g., Piner et al.\ 2009; Blasi et al.\ 2013).
Such limb brightening is important because it can reveal the 
presence of transverse velocity and/or magnetic field structures.
We note here that we continue to observe the presence of limb brightening in those two sources
in later epochs, and we also observe limb brightening in some of the new images
of other sources; for example 0645+153 at 2014 August 21 and 0706+592 at 2014 March 27 (see Figure~1).
We do not pursue this analysis further in this paper, but future work will investigate
the transverse jet structures measured from both the individual and the stacked-epoch images of all of the
sources in this program.

We also noted in Paper~I that two sources (0033+595 and 0647+251) showed jet components that differed by more than
$90\arcdeg$ in their position angles. Because of the likely at least modest Doppler factor
of the radio emission in these sources, such structure is unlikely to represent a jet and counter-jet.
Presumably then, either the brightest component is not the core
(in which case the component designated as component~2 in each source would be the core), or
the jet has a very large apparent bend (as inferred for the TeV blazar 1ES~1959+650 by Piner et al.\ 2008).
Images of these two sources were obtained at 15~GHz to compare with the 8~GHz images,
and these are also shown in Figure~1; in each case
the spectral index of the presumed core and of component~2 are similar within the estimated errors (see Table~\ref{mfittab}). 
However, we are confident in the core identification in each source due to the measured brightness temperatures, which are or order
$10^{10}$~K for the presumed core at every epoch, but only of order $10^{8}$~K for the closest jet component at
every epoch (for both sources); these can be compared with typical core brightness temperatures from Figure~4 of Paper~I.
We expect then that these two sources are cases where the jet has a large apparent bend due to projection effects; deep images
at lower frequencies would also be useful to confirm this.

\subsection{Model Fits}
\label{mfits}
In order to identify jet features from epoch to epoch,
we fit Gaussian models to the calibrated visibilities for each image in Figure~1, using the
{\em modelfit} task in DIFMAP. 
Model fitting directly to the visibilities rather than the images allows sub-beam
resolution to be obtained in many cases, and components may be clearly identified in the model fitting even when they
appear blended with the core component or with each other in the CLEAN images.
In some cases, patchy and low surface brightness emission beyond the collimated jet region
could not be well fit by Gaussian components,
so the model fits do not necessarily represent the more distant emission seen on the full CLEAN images.
Gaussian components will also not fully represent more complex transverse jet structure, such as limb-brightening.
Note also that, because of incomplete sampling in the $(u,v)$-plane,
such VLBI model fits are not unique, and represent
only one mathematically consistent deconvolution of the source structure.

During the model fitting, circular Gaussians were preferred to elliptical Gaussians if 
they provided an adequate fit to the visibilities, because their fit parameters are more stable
from epoch to epoch. An adequate fit was judged based on
the reduced chi-squared of the fit and visual inspection of the residual map and visibilities.
Elliptical Gaussians were used in the end only for two components: the core and the outermost jet component
of the source 0502+675. Additionally, if the size of a circular Gaussian component asymptotically approached zero during the 
model-fitting procedure, then that component was replaced with a delta function.

The Gaussian models fit to all 88 images are given in Table~\ref{mfittab}.
The model component identification follows the scheme used in our previous papers (e.g., Piner et al.\ 2010);
jet components are numbered 1, 2, etc., from the outermost component inward.
Component `0' indicates the presumed core at each epoch, while
a component ID of `99' indicates a flagged component not used in the analysis
(e.g., because it is a merger of two other components, or it is a more distant component not seen at other epochs;
such a component ID is assigned only twice for the 289 components in Table~\ref{mfittab}).
The polar coordinates of the center of each component in Table~\ref{mfittab} are relative
to the origin of the associated image in Figure~1, not relative to the core
(however, in most cases, the core position and the origin of the image are very close together).
Note that flux density values for closely spaced components in Table~\ref{mfittab} may be inaccurate, since it is difficult for
the fitting algorithm to uniquely distribute such flux density during the model fitting.

\begin{table*}[!t] 
\begin{center}
\caption{Gaussian Models}
\label{mfittab}
\begin{tabular}{l c c c c r r r c c c} \tableline \tableline \\ [-15pt]
Source & Epoch & Freq. & Comp. & \multicolumn{1}{c}{$S$} & \multicolumn{1}{c}{$r$} &
\multicolumn{1}{c}{P.A.} & \multicolumn{1}{c}{$a$} & $(b/a)$ & P.A.$_{\mathrm{maj}}$ & Type \\ [-3pt]
& & (GHz) & & \multicolumn{1}{c}{(mJy)} & \multicolumn{1}{c}{(mas)} &
\multicolumn{1}{c}{(deg)} & \multicolumn{1}{c}{(mas)} & & (deg) & \\ [-3pt]
(1) & (2) & (3) & (4) & \multicolumn{1}{c}{(5)} & \multicolumn{1}{c}{(6)} &
\multicolumn{1}{c}{(7)} & \multicolumn{1}{c}{(8)} & (9) & (10) & (11) \\ \tableline \\ [-15pt]
0011$-$191 & 2013.62 & 8.4 & 0 &  9.3 & 0.095 & 142.5   & 0.227 & 1.000 & 0.0 & 1 \\ [-3pt]
0011$-$191 & 2013.62 & 8.4 & 1 &  5.0 & 0.900 & $-$43.0 & 0.979 & 1.000 & 0.0 & 1 \\ [-3pt]
0011$-$191 & 2014.71 & 8.4 & 0 & 10.0 & 0.082 & 148.3   & 0.187 & 1.000 & 0.0 & 1 \\ [-3pt]
0011$-$191 & 2014.71 & 8.4 & 1 &  4.2 & 0.839 & $-$45.4 & 1.015 & 1.000 & 0.0 & 1 \\ [-3pt]
0011$-$191 & 2015.39 & 8.4 & 0 &  9.2 & 0.055 & 116.1   & 0.047 & 1.000 & 0.0 & 1 \\ [-3pt]
0011$-$191 & 2015.39 & 8.4 & 1 &  4.4 & 0.745 & $-$41.6 & 0.801 & 1.000 & 0.0 & 1 \\ [-3pt]
0011$-$191 & 2015.89 & 8.4 & 0 &  8.4 & 0.111 & 144.1   & 0.416 & 1.000 & 0.0 & 1 \\ [-3pt]
0011$-$191 & 2015.89 & 8.4 & 1 &  3.0 & 0.819 & $-$31.1 & 1.098 & 1.000 & 0.0 & 1 \\ \tableline \\ [-20pt]
\end{tabular}
\end{center}
\tablecomments{
Column~4: component identification. Component `0' indicates the presumed core. Other components are numbered from 1 to 5,
from the outermost component inward. A component ID of `99' indicates a flagged component not used in the analysis.
Column~5: flux density of the component in millijanskys.
Columns~6 and 7: $r$ and P.A. (position angle) are the polar coordinates of the
center of the component relative to the origin of the image in Figure~1 (not relative to the core).
Position angle is measured from north through east.
Columns~8--10: $a$ and $b$ are the FWHM of the major and minor axes of the Gaussian,
and P.A.$_{\mathrm{maj}}$ is the position angle of the major axis.
$(b/a)$ and P.A.$_{\mathrm{maj}}$ are set to 1.0 and 0.0 for circular components, respectively.
Column~11: component type. Type 1 indicates a Gaussian component, while
type 0 indicates a delta function.
\\ (Table~\ref{mfittab} is published in its entirety in machine-readable format.
A portion is shown here for guidance regarding its form and content.)}
\end{table*}

The number of components in the model fitting was purposely kept small enough for each source that components could be
easily identified from epoch to epoch across the full series of images. All series of model fits
for all sources have been verified to be temporally consistent with each other using the following procedure:
\begin{enumerate}\addtolength{\itemsep}{-.50\baselineskip}
\item{When the model fit for the first epoch is used as the starting guess for the second epoch and the iterative model fitting
proceeds, then the best fit for the second epoch given in Table~\ref{mfittab} is obtained.}
\item{This procedure can be repeated epoch by epoch until the best fit for the final epoch given in Table~\ref{mfittab} is obtained.}
\item{When the model fit for the final epoch is then used as the starting guess for the next-to-last epoch and the iterative model fitting
proceeds, then the best fit for the next-to-last epoch is obtained.}
\item{This procedure is repeated epoch by epoch until the original best fit for the first epoch is again obtained at the end.}
\end{enumerate}
Increasing the complexity of the model fits by adding too many components can disrupt this consistency and make it more
difficult to identify features between epochs. Because of the desire to obtain consistent sets of models over
all epochs, four of the twenty model fits published in Paper~I have been re-done for this paper.
The model fits for the sources 0031$-$196, 0150+015, 0502+675, and 2247+381 are thus slightly different here compared to
their corresponding model fits from Paper~I, while the other sixteen model fits from Paper~I remain identical in this paper.

\subsection{Apparent Speeds}
\label{speeds}

\begin{table*}
\begin{center}
\caption{Apparent Component Speeds}
\label{speedtab}
\begin{tabular}{l c r r r r c c} \tableline \tableline \\ [-15pt]
Source & Comp. & \multicolumn{1}{c}{$\langle{S}\rangle$} & \multicolumn{1}{c}{$\langle{r}\rangle$} & \multicolumn{1}{c}{$\mu$} 
& \multicolumn{1}{c}{$\beta_\mathrm{app}$} & $1\sigma$ & $2\sigma$ \\ [-3pt]
& & \multicolumn{1}{c}{(mJy)} & \multicolumn{1}{c}{(mas)} & \multicolumn{1}{c}{(mas yr$^{-1}$)} & & limit & limit \\ [-3pt]
(1) & (2) & \multicolumn{1}{c}{(3)} & \multicolumn{1}{c}{(4)} & \multicolumn{1}{c}{(5)} &
\multicolumn{1}{c}{(6)} & (7) & (8) \\ \tableline \\[-15pt]
0011$-$191 & 1 & 4.2  & $0.91\pm0.04$  & $-0.051\pm0.048$ & $-0.32\pm0.30$\tablenotemark{a} & ...  & ...  \\ [-3pt]
0031$-$196 & 1 & 4.7  & $3.42\pm0.06$  & $0.204\pm0.071$  & $6.17\pm2.14$\tablenotemark{a}  & 4.03 & 1.89 \\ [-3pt]
0033+595   & 1 & 6.1  & $5.49\pm0.10$  & $0.036\pm0.120$  & $1.00\pm3.39$ ~                 & ...  & ...  \\ [-3pt]
           & 2 & 9.8  & $1.18\pm0.05$  & $0.041\pm0.055$  & $1.16\pm1.54$\tablenotemark{a}  & ...  & ...  \\ [-3pt]
0133+388   & 2 & 2.7  & $0.94\pm0.10$  & $0.074\pm0.125$  & $1.82\pm3.05$ ~                 & ...  & ...  \\ [-3pt]
           & 3 & 5.5  & $0.45\pm0.04$  & $0.040\pm0.046$  & $0.98\pm1.13$\tablenotemark{a}  & ...  & ...  \\ [-3pt]
0150+015   & 1 & 1.0  & $7.34\pm0.39$  & $0.311\pm0.221$  & $1.64\pm1.17$\tablenotemark{a}  & 0.47 & ...  \\ [-3pt]
           & 2 & 2.8  & $2.94\pm0.11$  & $-0.119\pm0.064$ & $-0.63\pm0.34$ ~                & ...  & ...  \\ [-3pt]
           & 3 & 5.2  & $0.89\pm0.06$  & $-0.055\pm0.032$ & $-0.29\pm0.17$ ~                & ...  & ...  \\ [-3pt]
0229+200   & 1 & 3.0  & $15.81\pm0.28$ & $0.312\pm0.158$  & $2.85\pm1.44$ ~                 & 1.41 & ...  \\ [-3pt]
           & 2 & 1.1  & $6.64\pm0.05$  & $-0.062\pm0.028$ & $-0.56\pm0.25$ ~                & ...  & ...  \\ [-3pt]
           & 3 & 1.9  & $3.01\pm0.01$  & $0.059\pm0.005$  & $0.54\pm0.04$\tablenotemark{a}  & 0.50 & 0.46 \\ [-3pt]
           & 4 & 2.1  & $0.98\pm0.03$  & $-0.001\pm0.017$ & $-0.01\pm0.16$ ~                & ...  & ...  \\ [-3pt]
0317+185   & 1 & 1.0  & $5.39\pm0.05$  & $0.493\pm0.100$  & $6.04\pm1.23$\tablenotemark{a}  & 4.81 & 3.58 \\ [-3pt]
           & 2 & 1.4  & $2.14\pm0.11$  & $-0.017\pm0.219$ & $-0.21\pm2.68$ ~                & ...  & ...  \\ [-3pt]
           & 3 & 2.5  & $0.85\pm0.10$  & $0.018\pm0.073$  & $0.22\pm0.90$ ~                 & ...  & ...  \\ [-3pt]
0347$-$121 & 1 & 1.5  & $2.06\pm0.18$  & $0.136\pm0.102$  & $1.65\pm1.24$\tablenotemark{a}  & 0.41 & ...  \\ [-3pt]
0414+009   & 1 & 11.2 & $1.33\pm0.05$  & $0.002\pm0.032$  & $0.03\pm0.58$\tablenotemark{a}  & ...  & ...  \\ [-3pt]
0502+675   & 1 & 2.9  & $5.88\pm0.10$  & $0.375\pm0.166$  & $7.92\pm3.51$ ~                 & 4.41 & 0.90 \\ [-3pt]
           & 2 & 2.0  & $0.33\pm0.01$  & $0.105\pm0.021$  & $2.23\pm0.44$\tablenotemark{a}  & 1.79 & 1.35 \\ [-3pt]
0548$-$322 & 1 & 6.5  & $1.10\pm0.06$  & $-0.139\pm0.093$ & $-0.63\pm0.42$\tablenotemark{a} & ...  & ...  \\ [-3pt]
0645+153   & 1 & 3.0  & $21.97\pm0.10$ & $0.511\pm0.145$  & $5.90\pm1.67$\tablenotemark{a}  & 4.23 & 2.56 \\ [-3pt]
           & 2 & 1.8  & $10.65\pm0.20$ & $-0.018\pm0.301$ & $-0.21\pm3.48$ ~                & ...  & ...  \\ [-3pt]
           & 3 & 1.1  & $4.37\pm0.26$  & $0.021\pm0.386$  & $0.25\pm4.47$ ~                 & ...  & ...  \\ [-3pt]
           & 4 & 1.8  & $2.39\pm0.09$  & $0.069\pm0.142$  & $0.80\pm1.64$ ~                 & ...  & ...  \\ [-3pt]
           & 5 & 3.6  & $0.82\pm0.06$  & $0.006\pm0.098$  & $0.07\pm1.13$ ~                 & ...  & ...  \\ [-3pt]
0647+251   & 2 & 7.5  & $0.85\pm0.04$  & $-0.115\pm0.061$ & $-3.13\pm1.67$\tablenotemark{a} & ...  & ...  \\ [-3pt]
0706+592   & 1 & 4.6  & $13.25\pm0.35$ & $0.709\pm0.189$  & $5.79\pm1.54$\tablenotemark{a}  & 4.25 & 2.71 \\ [-3pt]
           & 2 & 4.3  & $3.31\pm0.06$  & $0.086\pm0.034$  & $0.70\pm0.27$ ~                 & 0.43 & 0.16 \\ [-3pt]
           & 3 & 5.8  & $0.82\pm0.01$  & $0.076\pm0.008$  & $0.62\pm0.06$ ~                 & 0.56 & 0.50 \\ [-3pt]
1008$-$310 & 1 & 3.7  & $1.41\pm0.09$  & $0.459\pm0.143$  & $4.27\pm1.33$\tablenotemark{a}  & 2.94 & 1.61 \\ [-3pt]
1221+248   & 1 & 1.8  & $2.12\pm0.07$  & $0.037\pm0.101$  & $0.52\pm1.41$ ~                 & ...  & ...  \\ [-3pt]
           & 2 & 2.0  & $0.84\pm0.05$  & $0.034\pm0.069$  & $0.47\pm0.96$\tablenotemark{a}  & ...  & ...  \\ [-3pt]
1440+122   & 2 & 1.4  & $1.72\pm0.14$  & $-0.065\pm0.199$ & $-0.69\pm2.10$ ~                & ...  & ...  \\ [-3pt]
           & 3 & 2.8  & $0.54\pm0.04$  & $0.025\pm0.058$  & $0.26\pm0.61$\tablenotemark{a}  & ...  & ...  \\ [-3pt]
1722+119   & 1 & 4.2  & $3.32\pm0.12$  & $-0.197\pm0.172$ & $-4.15\pm3.65$ ~                & ...  & ...  \\ [-3pt]
           & 2 & 7.2  & $0.54\pm0.03$  & $-0.006\pm0.038$ & $-0.12\pm0.81$\tablenotemark{a} & ...  & ...  \\ [-3pt]
1741+196   & 1 & 7.9  & $5.57\pm0.14$  & $-0.287\pm0.199$ & $-1.59\pm1.10$ ~                & ...  & ...  \\ [-3pt]
           & 2 & 17.5 & $2.25\pm0.04$  & $0.136\pm0.051$  & $0.75\pm0.28$\tablenotemark{a}  & 0.47 & 0.19 \\ [-3pt]
           & 3 & 24.9 & $0.79\pm0.02$  & $0.054\pm0.031$  & $0.30\pm0.17$ ~                 & 0.13 & ...  \\ [-3pt]
2247+381   & 1 & 2.5  & $5.00\pm0.03$  & $0.470\pm0.047$  & $3.65\pm0.36$\tablenotemark{a}  & 3.29 & 2.93 \\ [-3pt]
           & 2 & 4.3  & $1.54\pm0.06$  & $0.469\pm0.089$  & $3.64\pm0.69$ ~                 & 2.95 & 2.26 \\ [-3pt]
           & 3 & 9.7  & $0.59\pm0.03$  & $0.127\pm0.036$  & $0.99\pm0.28$ ~                 & 0.71 & 0.43 \\ \tableline \\[-20pt]
\end{tabular}
\end{center}
\tablenotetext{a}{Speed used for the histogram in Figure~4; see text for explanation.}
\tablecomments{
Column~1: source name. Column~2: component ID. Column~3: mean flux density.
Column~4: mean separation
from core. Column~5: proper
motion. Column~6: apparent speed in units of the speed of light. Columns~7 and 8: 1$\sigma$ and 2$\sigma$ lower
limits on the apparent speed, if greater than zero, respectively.}
\end{table*}

In order to the study the motions of the jet components,
we made linear least-squares fits to the separation of component centers from the core versus time,
for all 45 jet components from Table~\ref{mfittab} that were observed at four or more epochs.
We used the method described by Homan et al.\ (2001) to determine
the error bars on the component positions, modified for linear fits from their original version
for quadratic fits. This method uses the scatter of component positions about the
fit to estimate errors on model component positions that are not known {\em a priori},
and it was used in our previous work on the kinematics of TeV HBLs (Piner et al.\ 2010),
and on the kinematics of sources from the Radio Reference Frame Image Database (Piner et al.\ 2007, 2012).
After this fitting, we excluded two relatively diffuse and distant components whose positions were so poorly constrained that
they had proper motion errors exceeding 0.4~mas~yr$^{-1}$ (component 1 from 0133+388 and component 1 from 1440+122).
The remaining 43 fits to component motions are shown in Figure~2, and are tabulated in Table~\ref{speedtab}.

\begin{figure*}[!t]
\centering
\includegraphics[scale=0.82]{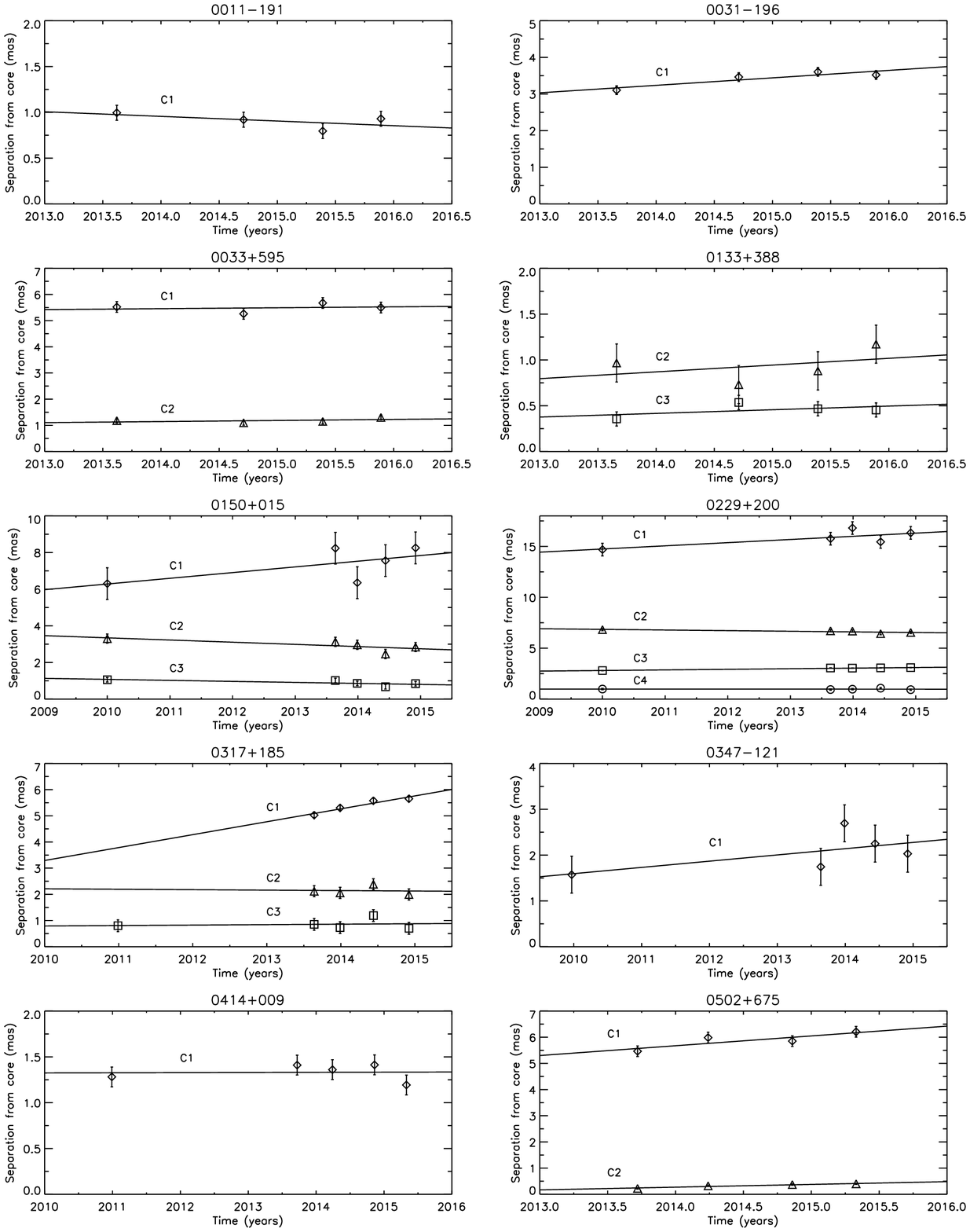}
\vspace{-0.30in}
\caption{Linear fits to the separation of model components from the core versus time, for
all components observed at four or more epochs.
Some error bars are smaller than the plotting symbols.}
\end{figure*}

\begin{figure*}
\centering
\includegraphics[scale=0.82]{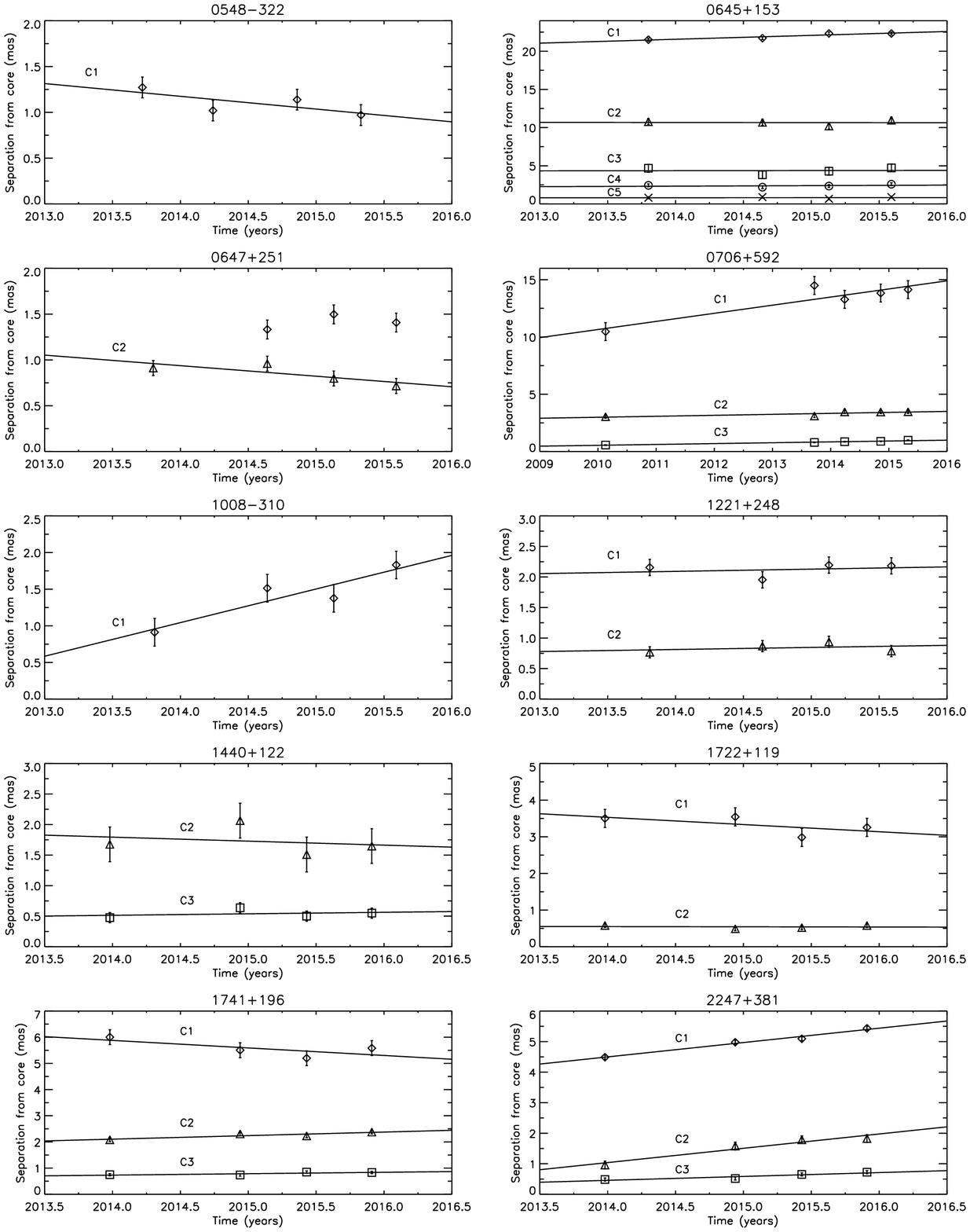}
\\Fig. 2.--—{\em Continued}
\end{figure*}

\begin{figure*}[!t]
\centering
\includegraphics[angle=90,scale=0.54]{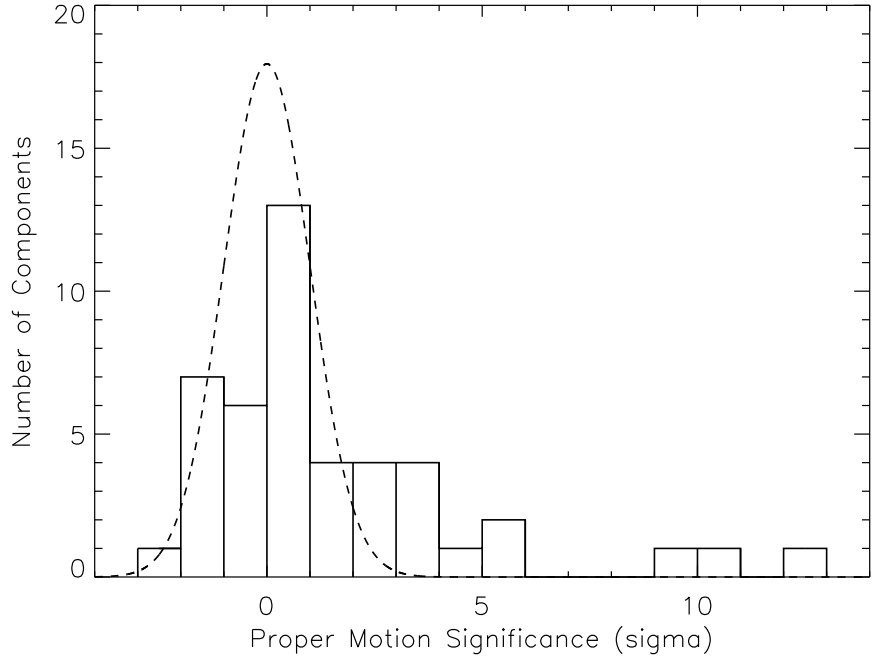}
\caption{Histogram of the significance of the proper motion measurement for each component,
for all 45 components that were observed at four or more epochs,
maintaining the negative sign for negative proper motion measurements. The dashed line is
the distribution expected for a population of stationary components.}
\end{figure*}

\begin{figure*}[!ht]
\centering
\includegraphics[angle=90,scale=0.54]{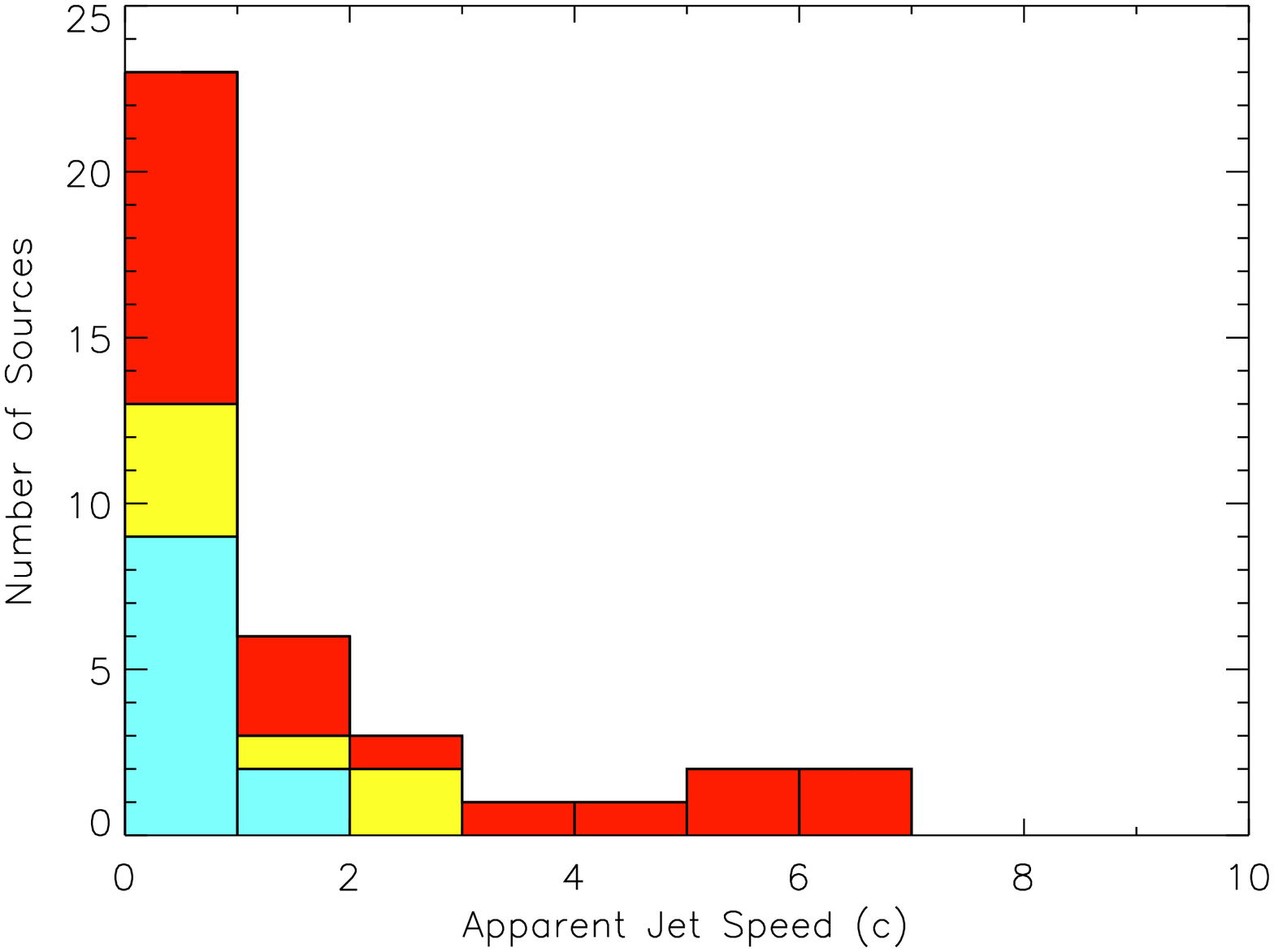}
\caption{Histogram of apparent speeds for the component in each source with the highest 
2$\sigma$ speed lower limit (or 1$\sigma$ if no component has 2$\sigma$ significance).
New sources with VLBI data from this paper are shown in red (20 sources),
sources with data taken from our previous papers (Piner et al.\ 2010; Tiet et al.\ 2012) are shown in blue (11 sources),
and sources with data taken from the MOJAVE program are shown in yellow (7 sources).}
\end{figure*}

For each of these 43 components, Table~\ref{speedtab} lists the average flux density of the component,
the average separation from the core obtained if a constant separation is fit to the component positions,
the proper motion obtained from the linear fit, and the apparent speed obtained from that proper motion using
the redshift given in Table~\ref{sourcetab}
(for the two sources in Table~2 with lower limits, we have adopted the lower limit as the redshift).
Average flux densities of the fitted jet components
range from 1.0 to 24.9~mJy, with a median flux density of 3.0~mJy.
The median positional error bar size for all 185 data points fit in Figure~2 is 0.13~mas,
which is about 10\% of the median beam size from Table~\ref{imtab}, so that on average jet components
are being localized to within about 1/10 of the naturally weighted beam.
Because most components were observed over a time baseline of about two years, we expect a
typical proper motion error of order 0.07 mas~yr$^{-1}$, which is indeed the median proper motion
error of the 43 fits in Table~\ref{speedtab}. At the median redshift of our sample ($z=0.18$), this proper motion
error translates into an apparent speed error of about $1c$, which is consistent with the median apparent speed
error from Table~\ref{speedtab}. These observations thus achieved their original goal of
constraining the apparent motions in these 20 HBL jets with an accuracy of order $\pm1c$ over about two years.

As has been found in earlier work on the TeV HBLs (e.g., Piner et al.\ 2010; Lico et al.\ 2012; Tiet et al.\ 2012), many of the components
whose proper motions are given in Table~\ref{speedtab} appear stationary within the measurement errors.
Nevertheless, a subset of significant outward apparent motions is detected, as can be seen from an analysis of the
proper motion significances. 
Figure~3 shows a histogram of the significance in multiples of sigma of the proper motion measurement 
for each component (maintaining the negative sign for negative proper motion measurements),
for all 45 components that were observed at four or more epochs.
The dashed curve in Figure~3 shows the theoretical distribution expected for a population of truly stationary components.
As can be seen from Figure~3, the negative proper motions that we have observed at between 1 and 3$\sigma$ significance
are consistent with the expected scatter about zero motion, so that we cannot claim any detection of significant
inward apparent motions. However, the distribution in Figure~3 has a bias toward
positive proper motions that significantly exceeds that expected for a population of stationary components.
For example, 14 out of 45 components (or about one third) have a positive proper motion that exceeds 2$\sigma$ significance,
where only about one is expected by chance.
Since outward motions in these jets are detected, we proceed with analyzing these motions in more detail in the next section.

\section{Analysis and Discussion}
\label{discussion}

\begin{figure*}[!t]
\centering
\includegraphics[angle=90,scale=0.54]{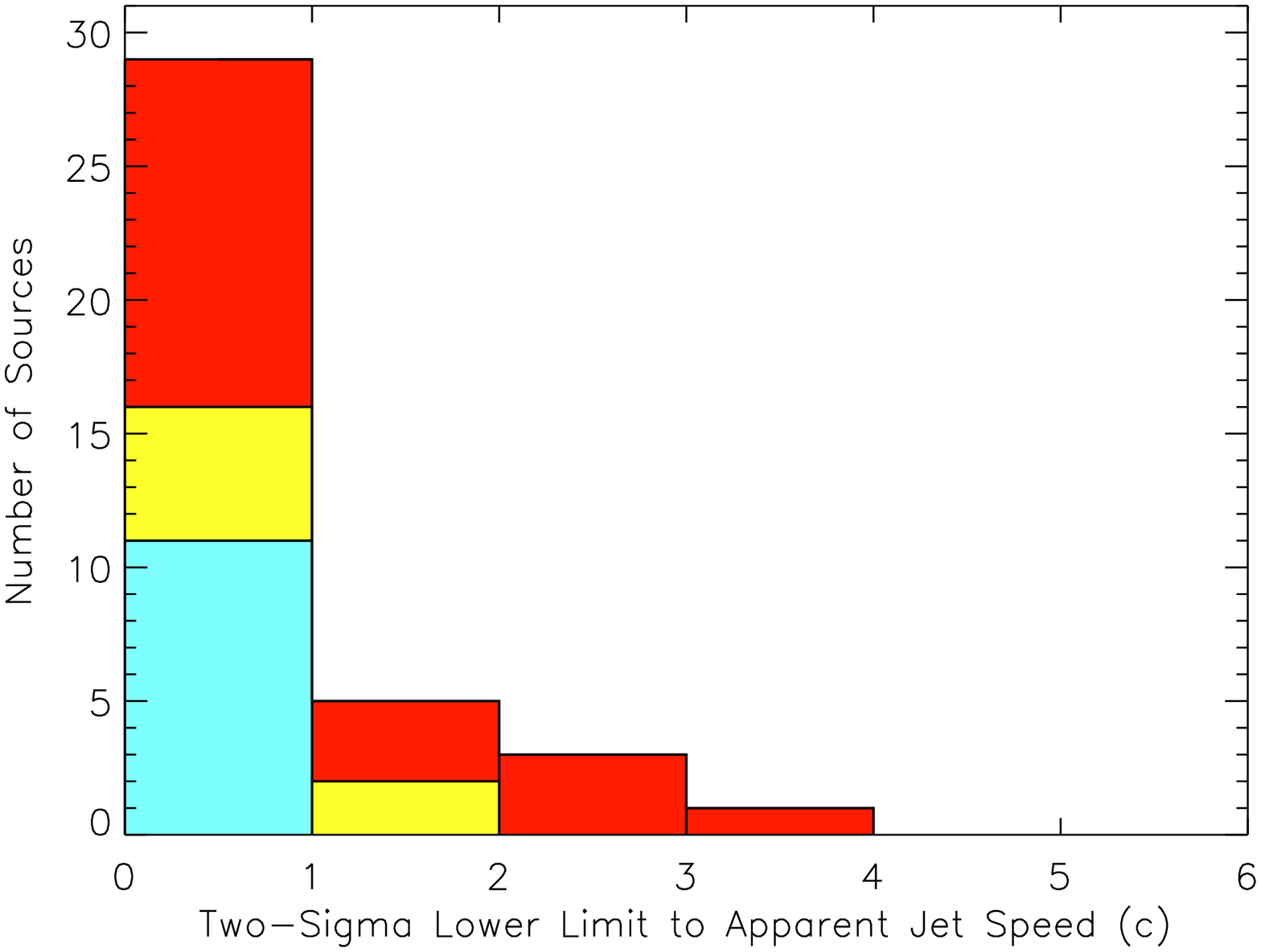}
\caption{Histogram of $2\sigma$ lower limits to the apparent speeds for the components plotted in Figure~4.
New sources with VLBI data from this paper are shown in red (20 sources),
sources with data taken from our previous papers (Piner et al.\ 2010; Tiet et al.\ 2012) are shown in blue (11 sources),
and sources with data taken from the MOJAVE program are shown in yellow (7 sources).}
\end{figure*}

In this section we relate the apparent speed measurements from the previous section 
to the bulk properties of the jets.
There are several effects that may produce pattern speeds of model components that are either stationary or slower than
the bulk apparent flow speed. These may include physical effects such as standing shocks or trailing features 
(e.g., Gomez et al.\ 1995; Kadler et al.\ 2008; Hervet et al.\ 2016, 2017), and modeling effects such as exceptionally
smooth flows without discernible local maxima to track. We note that while the jets of some of the TeV HBLs do appear smooth,
many do display local peaks in the jet that can be followed over time (see Piner \& Edwards 2004, and Figure~1 of this paper).
Cohen et al.\ (2014, 2015) also report that components in the jet of the IBL BL Lacertae may represent MHD waves
that move at apparent speeds exceeding the bulk apparent speed, although no such components have yet been reported in an HBL jet.
Because of these various effects, VLBI surveys have tended to use the fastest measured apparent speed in a jet as being the one that may be most indicative of
the peak bulk apparent speed of the flow (e.g., Lister et al.\ 2009b, 2013, 2015, 2016; Piner et al.\ 2012), 
and we have followed a similar practice in our previous work
on the kinematics of TeV HBLs (Piner et al.\ 2010; Tiet et al.\ 2012). We continue to follow such a practice for this
paper, with some necessary modifications as described below.

Because this work is focused on relatively newly discovered TeV blazars,
many of the sources analyzed for this paper have been monitored with VLBI for only about two years at 8~GHz, and thus
a number of the speed measurements have large associated errors. Because of the presence of potentially large associated errors, it is not useful
to simply use the fastest measured speed in a source regardless of the significance of the measurement.
Therefore, to assign an apparent speed to a source, we use the speed of the component that has the highest $2\sigma$ speed lower limit,
or, if no component in a source has at least $2\sigma$ significance, then we use the speed of the component that has the
highest $1\sigma$ speed lower limit. The histogram of these apparent speeds is shown in Figure~4, and the 
component used to plot each source in Figure~4 is indicated by a note in Table~\ref{speedtab}. 
This procedure has the effect of maximizing the tail of the distribution of the $2\sigma$ speed lower limit histogram,
which is important in the subsequent analysis.
Because we showed in the previous section that all formally negative apparent speeds are consistent
with a random scatter about zero, any such speeds are plotted in the left-most bin of Figure~4.

\begin{figure*}[!t]
\centering
\includegraphics[angle=90,scale=0.57]{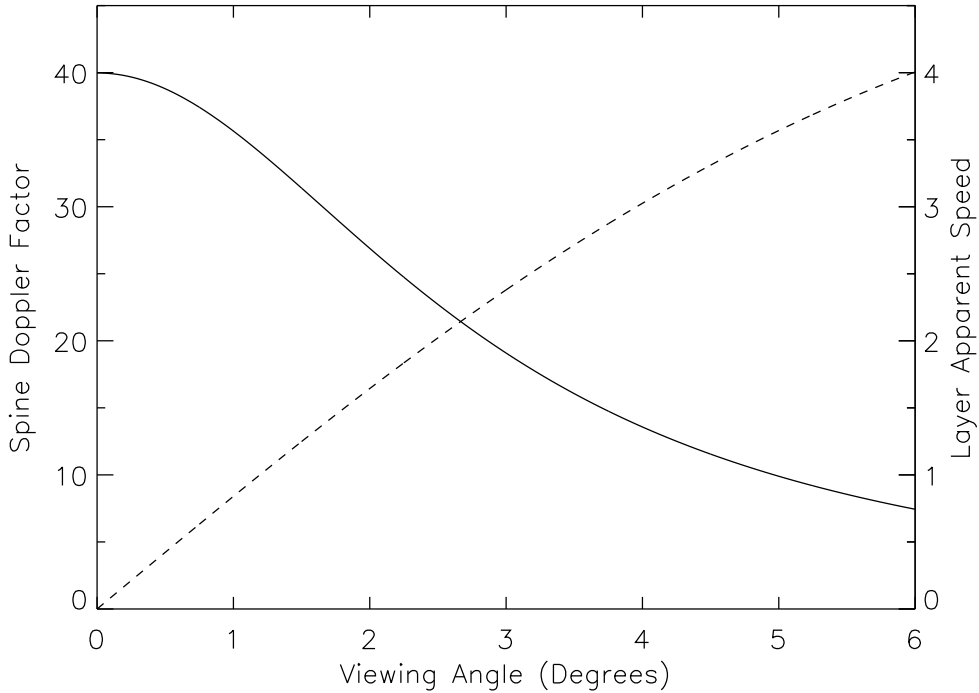}
\caption{The solid curve and left axis show the Doppler Factor of a jet spine with a Lorentz factor of 20 versus viewing angle of the jet.
The dashed curve and right axis show the apparent speed (in units of $c$) of a jet layer with a Lorentz factor of 5 versus the same jet viewing angles.} 
\end{figure*}

In Figure~4 we have also included measured apparent speeds for all other TeV HBLs for which they are available (assigned using the same procedure
described above). These additional sources are either compiled from our earlier work in
Piner et al.\ (2010) and Tiet et al.\ (2012) (11 sources), or from observations by the MOJAVE program
(7 sources). These are the 18 sources that are indicated by an exclusion code of either `1' or `2' in Table~\ref{selecttab}.
For all sources present both in our earlier work and in the MOJAVE program, the two datasets agree on the histogram bin.
For five of the seven sources taken from the MOJAVE program, we use an apparent speed from previously published works.
These apparent speeds are: $2.33\pm0.51c$ for PKS~0301$-$243 (Lister et al.\ 2016), $0.83\pm0.04c$ for IC~310 (Glawion et al.\ 2016),
$1.78\pm0.37c$ for 1ES~1011+496 (Lister et al.\ 2013), $0.032\pm0.014c$ for 1ES~1215+303 (Lister et al.\ 2013),
and $2.6\pm1.1c$ for PKS~1424+240 (Lister et al.\ 2013). For two additional sources from the MOJAVE program
without published apparent speeds (1ES~0806+524 and 1ES~1727+502), we have independently fit Gaussian
models to the publicly available visibility data, and then measured apparent speeds using the same procedure as for
our other sources (see $\S$~\ref{mfits} and \ref{speeds}). For 1ES~0806+524 we fit two components with apparent speeds of
$-0.02\pm0.18c$ and $0.05\pm0.18c$, and for 1ES~1727+502 we fit four components with apparent speeds of
$-0.58\pm0.29c$, $-0.04\pm0.26c$, $-0.06\pm0.13c$, and $-0.18\pm0.12c$ (so all consistent with no motion).
When all of these additional sources are combined with the 20 new sources from this paper, 
Figure~4 then includes all 38 of the 47 TeV HBLs for which multi-epoch structural information is available (see Table~\ref{selecttab}).
Inspection of Figure~4 shows that half of the newly added sources are subluminal, 
but that there is also a small tail present
that shows apparent speeds extending to above $3c$ for the first time. 

Below we first discuss the nature of the superluminal tail of the distribution, followed by the nature of the subluminal and stationary components.
The six components with apparent speeds above $3c$ are discussed individually in the notes on individual sources following this section. 

\subsection{Superluminal Components}
Because the few components that are in the higher speed tail of Figure~4 are likely to lie near the upper extent
of their allowed error range, we plot in Figure~5 a histogram of the $2\sigma$ lower limits to the
apparent speeds of all of the components plotted in Figure~4. The color scheme in Figure~5 has the same meaning as in the previous figure.
As in Figure~4, formally negative values are plotted in the left-most bin.
The distribution of lower limits extends out to about $4c$, and in fact
components were selected for plotting in Figure~4 in order to maximize the extent of the tail of the distribution
of $2\sigma$ lower limits in Figure~5.
Thus, even with fairly conservative $2\sigma$ lower limits, we find that when these new sources are included, that apparent speeds of at least
a few times the speed of light are observed in a small minority of the TeV HBLs.
However, according to Figure~5, no apparent speed significantly exceeding about $4c$ has been detected in any
of these sources throughout the history of the monitoring programs. This is a key observational result that in turn implies peak bulk Lorentz factors
of order 4~in the parsec-scale radio emitting portions of these jets. This rather low value for the bulk Lorentz factor is entirely consistent
with that found by other VLBI estimates, such as the radio core brightness temperatures (Paper~I;
Lister et al.\ 2011; Lico et al.\ 2016).

As has been found previously (Kharb et al.\ 2008; Piner et al.\ 2010; Tiet et al.\ 2012; Lister et al.\ 2016), 
we confirm that the apparent speed distribution of the TeV HBLs consists of
significantly lower apparent speeds than is found for other source classes. 
For comparison, the peak apparent speeds of other source classes, as measured by the MOJAVE survey, 
are about $50c$ for quasars, $20c$ for other BL Lac objects (LBLs and IBLs), and about
$10c$ for radio galaxies and radio-loud narrow-lined Seyfert I AGNs (see Figure~8 from Lister et al.\ 2016).
The HBLs thus comprise a kinematically distinct class compared to other AGNs with parsec-scale radio jets.

The implied peak Lorentz factors of a few from the radio observations of TeV HBL jets conflict with the high Doppler factor and Lorentz factor
estimates based on variability and SED modeling of the high-energy emission, a contradiction that has become
known as the `Doppler crisis' (see $\S$~\ref{intro}). The differing values of the Lorentz factor estimated using 
data from different portions of the electromagnetic spectrum has led to the idea of velocity structures in the jets of TeV HBLs.
One possible geometry for these velocity structures that is also physically motivated by both theoretical simulations
of jets and unification work is that of a fast central spine that dominates the high-energy emission, and
a slower outer layer that dominates the radio emission (e.g, Ghisellini et al.\ 2005).
Note that in such models the radio emission from the layer can exceed that of the spine even if the layer has
a lower Doppler factor, due to the differing SEDs and emissivities between the spine and layer;
see, for example, Ghisellini et al.\ (2005) and Sahayanathan (2009).
Here we show how such a spine-layer scenario might plausibly explain observational results such as
the apparent speed lower limit histogram in Figure~5.

Figure~6 shows the Doppler factor of the spine of a hypothetical spine-layer jet with a Lorentz factor of
20 as the viewing angle changes from zero to six degrees (solid line).
This figure also shows the apparent speed of a hypothetical layer with a Lorentz factor of 5
over the same range of angles (dashed line). As the Doppler factor of the spine falls from about 40 to about 10,
the apparent speed of the layer increases from zero to about $4c$.
Thus, if there is a span of about a factor of four in the Doppler factor of the TeV blazar population,
this could accommodate, with no other differences, the ranges of apparent speeds seen in Figure~5.
(A change in the Doppler factor by a factor of four changes the integrated TeV energy flux above
a common threshold energy by about two orders of magnitude, using typical spectral indices from Paper~I,
with the exact value depending on the geometry of the emitting region. This is roughly the spanned range
of the integrated TeV energy fluxes as computed from data in Paper~I, so such a change seems plausible.)
This would also explain why the tail of the distribution in Figure~5 is occupied by the more
recently detected sources: as fainter TeV sources are detected as telescope sensitivity improves, these may be revealing lower
Doppler factor spines seen at larger viewing angles, which in turn would result in increasing apparent speeds for the layer.
Consequently, apparent speeds of a few times the speed of light may be expected in a spine-layer scenario
as more sources with fainter spine emission are observed, although this
should be confirmed with more rigorous Monte Carlo simulations that take into
account a full range of variables (such as luminosity, redshift, and Lorentz factor distributions).
Note also that the analysis in Figure~6 is not limited to a spine-layer geometry, but could apply
to other suggested geometries for the TeV HBLs, as long as there are two regions of the flow with different Lorentz factors;
e.g., the non-steady magnetized outflow model by Lyutikov \& Lister (2010).

If TeV blazar jets are composed of a fast spine and a slower layer, then radiative interaction between these
two regions may serve to decelerate the spine while simultaneously accelerating the layer (Georganopoulos \& Kazanas 2003; Ghisellini et al.\ 2005).
Such radiative acceleration of the layer in the context of the spine-layer model has recently been
investigated theoretically by Chhotray et al.\ (2017).
Here we test to see if there are any systematic changes in apparent speed with distance from the core
at the length scales probed by our current data. We have not observed individual components at a large enough number of epochs
to reliably fit for accelerations of individual components as was done by, e.g., Homan et al.\ (2009, 2015) and Piner et al.\ (2012).
Instead, we use a different method also employed by Piner et al.\ (2012).
We perform a fit to $\ln\beta_\mathrm{app}$ versus $\ln\langle{r}\rangle$ for components where motion is detected at above the $1\sigma$ level, 
using the measured values from Table~\ref{speedtab},
for each of the five sources with at least two such $1\sigma$ components 
(0229+200, 0502+675, 0706+592, 1741+196, and 2247+381).
A constant positive apparent acceleration along the length of the jet would yield a slope of 0.5 for such a fit.
All five of these sources have a positive slope for this fit, and
the weighted mean slope is $0.60\pm0.07$, approximately consistent with constant acceleration.
The binomial probability of all five sources having positive slope by chance is 0.03, consistent with the marginal
detection of apparent acceleration at a significance of 0.97, although given the small number of sources used for the test,
it should be confirmed by future studies. We note that Lister et al.\ (2013) found a positive correlation
between apparent speed and distance from the core for the more powerful BL Lacs in the MOJAVE sample, so
that this kinematic property does not seem to be unique to the HBL class, leaving open the question of whether
in the TeV HBLs it is due to the putative spine-layer interaction, or due to some other more general effect.

\subsection{Subluminal and Stationary Components}
The nature of the model components whose fitted speeds are consistent with no motion is likely to be some mix
of the following two cases:
\begin{enumerate}\addtolength{\itemsep}{-.50\baselineskip}
\item{Stationary or slowly moving patterns that do not reflect the bulk apparent speed of the underlying flow
(this is likely the case in sources where such components co-exist with much more rapidly moving components
such as in 0645+153)} 
\item{Components moving at the bulk apparent speed in jets that truly do have a 
slow apparent bulk speed (for example, jet layers at small angles to the line of sight,
such as those on the small viewing angle side of Figure~6).}
\end{enumerate}
Related to the first case above, Hervet et al.\ (2016, 2017) have argued
that stationary patterns due to recollimation shocks may be more common in HBLs than they are
in other types of blazars, and our observations of numerous apparently stationary components seem to be consistent with this. 
However, as we have shown in Figures~3 and 5, based both upon our data and the included MOJAVE data, 
a minority population of moving components is also present. 
We also note that a few of the components in Table~\ref{speedtab} that have measured subluminal speeds
are moving outward with high significance, just subluminally (e.g., component 3~in 0229+200),
and that these are likely examples of the second case above. For any specific jet that consists solely
of components that are consistent with no motion (e.g., 1221+248), the difference between the two
cases described above would be difficult to determine from these data alone. However, for slow bulk apparent speeds (rather than stationary patterns), 
motion should become measurable at some level if the VLBI monitoring data become more extensive. 

A self-consistent explanation for why the TeV HBLs apparently have jets with velocity structures that
observationally lead to the `Doppler crisis' may be something like the following.
A TeV-selected sample selects low-luminosity sources: either because this selection favors rare high-synchrotron peak sources which are drawn
from the low end of the luminosity function where the source density is largest (Giommi et al.\ 2012),
or because spectral peak frequencies are anti-correlated with luminosity in a blazar sequence (e.g., Ghisellini et al.\ 2017).
At this low end of the luminosity function the jets are formed in a low-efficiency accretion mode
(Ghisellini et al.\ 2005, 2009; Meyer et al.\ 2011; Sbarrato et al.\ 2014) that favors interaction of
the jet walls with the external medium, causing the formation of a slower layer (e.g., Rossi et al.\ 2008).
These flows may favor the generation of stationary recollimation shocks
that help to contribute to the large population of subluminal components in Figures~4 and 5 (Hervet et al.\ 2016, 2017).
Interaction between the spine and the layer may then decelerate the spine (e.g., Georganopoulos \& Kazanas 2003; Ghisellini et al.\ 2005; Chhotray et al.\ 2017),
producing longitudinal as well as transverse velocity structures.
Adoption of multi-zone geometries can then somewhat reduce the discrepancy in the Doppler factors that originally
led to the `Doppler crisis', because lower Doppler factors are typically needed to model the high-energy
emission when compared to single-zone models (due to radiative interaction between the two regions; e.g., Aleksi{\'c}, et al.\ 2014).
The picture described above has emerged based on theoretical modeling, high-resolution imaging, and unification studies,
and further work in all three areas should help to test and refine it.

It is also interesting to consider these results from the perspective of the unification of low-luminosity radio sources.
At lower accretion rates the FR~I LERGs may transition to make up a portion of the so-called `FR~0' population
(see, e.g., Figure~8 of Ghisellini 2011) that has been revealed in surveys of nearby radio sources (e.g., Baldi \& Capetti 2009; Sadler et al. 2014).
Such objects appear to lack extended radio emission, and most show no evidence of relativistic beaming (Sadler et al. 2014).
Some of these objects may represent weak jets that are vulnerable to instabilities and disruption, and that are unable to develop larger radio structures.
Observations of the `Doppler crisis' in the HBL jets may be showing the development of such instabilities.
VLBI studies of these `FR~0' sources will be important to conduct in this context, in order to see how their parsec-scale radio properties
compare to those of the HBL sources studied here.

\begin{figure*}[!t]
\centering
\includegraphics[angle=90,scale=0.55]{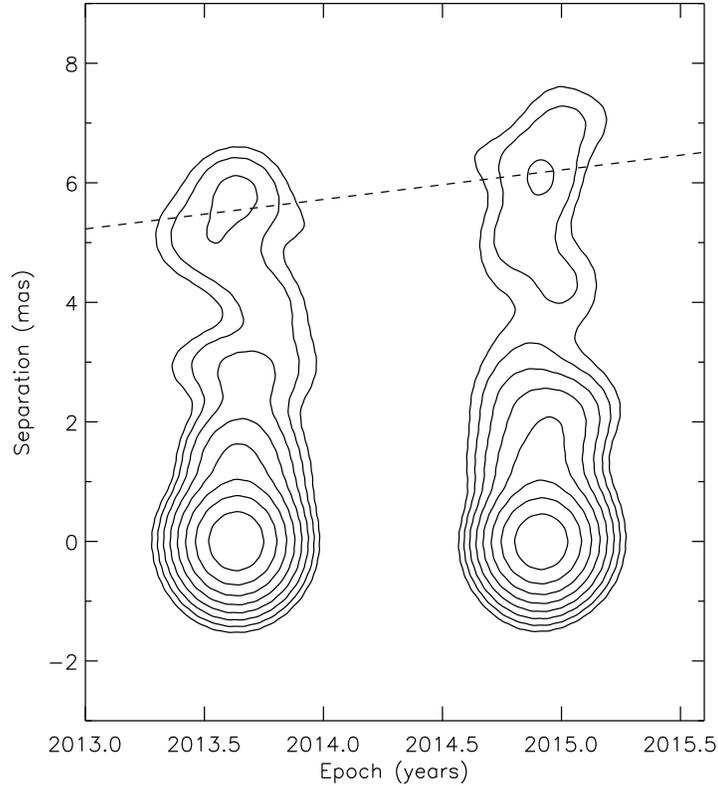}
\caption{Images of 0317+185 at the second and fifth epochs. The dashed line shows motion at the fitted speed of component~1.
Images are restored with a circular 1~mas beam. The lowest contour is 0.075~mJy~bm$^{-1}$, subsequent contours are
factors of two higher.}
\end{figure*}

\begin{figure*}[!t]
\centering
\includegraphics[angle=90,scale=0.55]{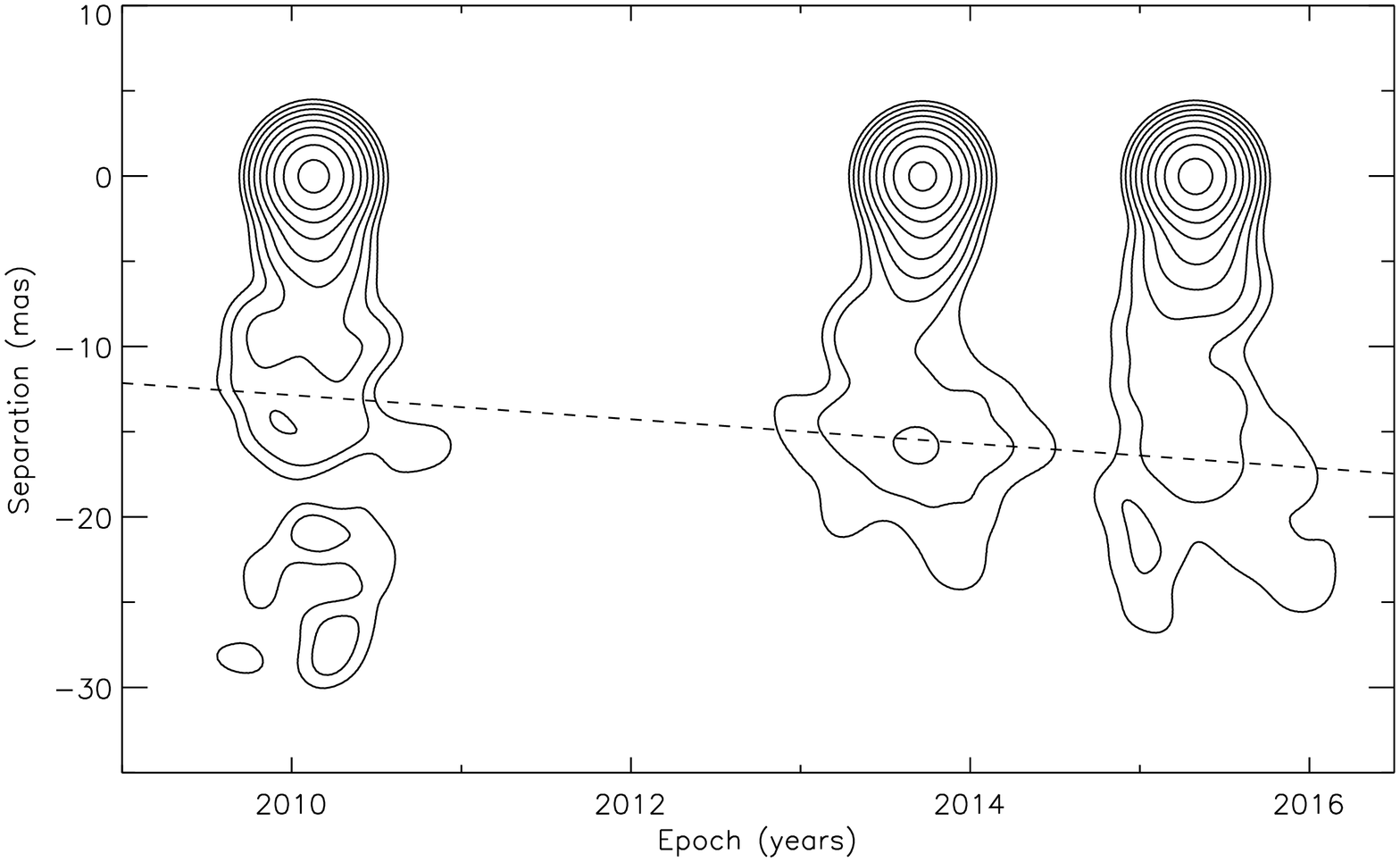}
\caption{Images of 0706+592 at the first, third, and fifth epochs. The dashed line shows motion at the fitted speed of component~1.
A taper has been applied to the visibility data, and images are restored with a circular 3~mas beam. 
The lowest contour is 0.1~mJy~bm$^{-1}$, subsequent contours are
factors of two higher.}
\end{figure*}

\section{Notes on Individual Sources}
Because of their importance in the analysis above, we focus here on the quality of the apparent speed 
measurement for each of the six components whose measured apparent speed is $>3c$ (see Figures~4 and 5).
We assign a quality code
to each of these component motions using the criteria developed by Kellermann et al.\ (2004)
for the 2~cm VLBA survey. These criteria were designed for an older VLBI survey preceding the MOJAVE survey, but this 
means that they were also designed to be applicable to a VLBI survey with a small and limited
number of epochs such as the one described here, and where the only goal is the measurement of component apparent speeds,
and not anything more sophisticated such as acceleration. These criteria are:
\begin{enumerate}\addtolength{\itemsep}{-.50\baselineskip}
\item{The component is observed at four or more epochs. This applies to all components in Table~\ref{speedtab}.}
\item{The component is a well-defined feature in the images. Note that in some cases a component may
be a well-defined feature in images made with tapered visibility data, but it is resolved out into patchy emission in
the full-resolution images in Figure~1.}
\item{The uncertainty in the fitted proper motion is $\leq0.08$ mas yr$^{-1}$, or the proper motion has a significance
$\geq5\sigma$.}
\end{enumerate}
The quality codes are then assigned as follows:
\begin{enumerate}\addtolength{\itemsep}{-.50\baselineskip}
\item{`Excellent' for motions that satisfy all three of the above criteria.}
\item{`Good' for motions that satisfy any two of the above criteria.}
\item{`Fair' for motions that satisfy only one of the above criteria.}
\item{`Poor' for motions that do not satisfy any of the above criteria, or for motions where
the uncertainty in the fitted proper motion is $>0.15$ mas yr$^{-1}$ (except for the $\geq5\sigma$
cases mentioned above).}
\end{enumerate}
Applying these criteria to the six components whose measured apparent speed is $>3c$, we obtain the following:
\begin{description}\addtolength{\itemsep}{-.50\baselineskip}
\item[0031$-$196] Component~1 is observed at four epochs, with a proper motion of $0.204\pm0.071$ mas yr$^{-1}$.
It is seen as a `shoulder' of emission off of the core in some images, so is not always a well-defined feature.
It receives a quality code of `Good' using the system above.
\item[0317+185] Component~1 is observed at four epochs, with a proper motion of $0.493\pm0.100$ mas yr$^{-1}$.
It is a well-defined feature whose motion is clearly evident, as is shown in the two-epoch image comparison in Figure 7.
It receives a quality code of `Good'.
\item[0645+153] Component~1 is observed at four epochs, with a proper motion of $0.511\pm0.145$ mas yr$^{-1}$.
It is fairly distant from the core, and is over-resolved in the images in Figure~1. However, it is seen as a
distinct feature in tapered images (see the following source for an example of this). It has a quality code of `Good'.
\item[0706+592] Component~1 is observed at five epochs, with a proper motion of $0.709\pm0.189$ mas yr$^{-1}$.
It is seen as a distinct feature in tapered images of this source, as is shown in the three-epoch image comparison in Figure 8.
However, the relatively large uncertainty in the proper motion gives it a `Poor' quality code.
\item[1008$-$310] Component~1 is observed at four epochs, with a proper motion of $0.459\pm0.143$ mas yr$^{-1}$.
It is a `shoulder' of emission off of the core rather than a well-defined feature. It has a quality code of `Fair'.
\item[2247+381] Component~1 is observed at four epochs, with a proper motion of $0.470\pm0.047$ mas yr$^{-1}$.
It is a well-defined feature, and has an `Excellent' quality code.
\end{description}
Thus, four of the six relatively fast components have quality codes of `Excellent' or `Good' in this system, and four
of the six motions are easily visible in either the full resolution or the tapered images.

\section{Conclusions}
\label{conclusions}
The HBLs are a physically important class of sources that
possess parsec-scale jet kinematics that is clearly distinct from the more powerful sources; however,
because of their relative faintness in the radio, they have not been previously well-studied, apart from a handful
of sources. In this paper, we have presented parsec-scale apparent speed measurements for 20 new TeV HBLs, based on 88 multi-epoch VLBA images.
These measurements were combined with data on 18 other sources from the literature to provide parsec-scale jet kinematics
for 38 of the 47 known TeV HBLs. To our knowledge, this is the largest published set of kinematic information on the HBL source class. 
Importantly, our sample has imposed no radio flux density limit
(in contrast to the 0.1~Jy flux density limit of the MOJAVE survey, for example (Lister et al.\ 2016)), but has observed all TeV-detected HBLs
regardless of their faintness in the radio.

In agreement with earlier works, we have confirmed that the measured apparent speeds
of jet components in the TeV HBLs are considerably slower than in the more powerful sources.
Many of these jet components have measured apparent speeds that are consistent with no motion,
and some of these may represent stationary patterns. A small number of mildly
superluminal components is detected in the TeV HBLs for the first time; the highest 
2$\sigma$ apparent speed lower limit considering all of the monitored TeV HBLs
from this paper is $3.6c$. No component with an apparent speed lower limit exceeding this 
has been detected by us, despite the fact that nearly all known
TeV HBLs have now been monitored with VLBI. This suggests that bulk Lorentz factors of up to about 4, but probably
not much higher, exist in the parsec-scale radio emitting regions of these sources,
consistent with estimates obtained in the radio by other means such as brightness temperatures.

Such Lorentz factors are reconciled with the high Lorentz factors obtained
at other wavebands by inferring that these jets contain different emission regions
with different Lorentz factors. A jet with a fast inner spine and slower outer layer
represents one such structure that is also expected based on theoretical grounds
and from arguments based on radio source unification. Our apparent speed results
may represent a population of such jets where the spine Doppler factor decreases
and the layer apparent speed increases as sources are observed at larger angles to
the line of sight.

Future work to be undertaken on the set of TeV HBLs described in this paper includes stacking of the VLBA images over
all available epochs to investigate fainter jet structures, including possible transverse structures;
monitoring of the five TeV HBLs that were detected too recently to be included in this set of multi-epoch monitoring (see Table~\ref{selecttab});
and continued monitoring of the six highest apparent speed sources from the current program (see Figure~4). The increased number of
epochs on those kinematically interesting sources will allow a better investigation of any apparent accelerations in the jets.

The set of TeV HBLs that is described in this paper is likely to be close to the complete set of these objects
that will be detected by the current generation of TeV telescopes, because the most promising candidate sources have
now been observed, and the number of new detections has been steadily declining. For example, according to TeVCat, while five new TeV HBL detections
were announced during the two years 2013-2014, only one was announced during the following two years 2015-2016.
The start of observations of the Cherenkov Telescope Array (CTA)
in about 2021 should reveal many more examples of this class of object. The potential faintness of these new objects in the radio
may pose an interesting challenge for VLBI imaging, although successfully imaging the CTA detections on the parsec scale will be
crucial to understanding the geometry and physics of their high-energy emission regions.

\acknowledgments
The Long Baseline Observatory is a facility of the National Science Foundation 
operated under cooperative agreement by Associated Universities, Inc.
This research has made use the TeVCat online source catalog (http://tevcat.uchicago.edu). 
This research has made use of the NASA/IPAC Extragalactic Database (NED) 
which is operated by the Jet Propulsion Laboratory, California Institute of Technology, 
under contract with the National Aeronautics and Space Administration.
This research has made use of data from the MOJAVE database that is maintained by the MOJAVE team (Lister et al.\ 2009a).
Part of this research was carried out at the Jet Propulsion Laboratory,
Caltech, under a contract with the National Aeronautics and Space Administration.
This work was supported by Fermi Guest Investigator grant NNX15AB93G.
This research has made use of NASA's Astrophysics Data System.

\facility{VLBA ()}

\end{document}